\documentclass[useAMS,usenatbib]{mn2e}

\usepackage{graphicx}
\usepackage{aas_macros}
\usepackage{placeins}
\usepackage{graphicx}    % needed for including graphics e.g. EPS, PS

\title[Connecting the dots II]{Connecting the dots II: Phase changes in the climate dynamics
of tidally locked terrestrial exoplanets}
\author[L. Carone, R. Keppens and L. Decin]{L. Carone$^{1}$\thanks{E-mail:
ludmila.carone@wis.kuleuven.be (LC)},  R.
Keppens$^{1}$, and L. Decin$^{2}$\\
$^{1}$Centre for mathematical Plasma Astrophysics, Department of Mathematics, KU Leuven, Celestijnenlaan 200B, 3001 Leuven, Belgium\\
$^{2}$Instituut voor Sterrenkunde, KU Leuven, Celestijnenlaan 200D, 3001 Leuven, Belgium}
\begin{document}

\date{TBD}

\pagerange{\pageref{firstpage}--\pageref{lastpage}} \pubyear{TBD}

\maketitle

\label{firstpage}

\begin{abstract}
We investigate 3D atmosphere dynamics for tidally locked terrestrial planets with an Earth-like atmosphere and irradiation for different rotation periods ($P_{rot}=1-100$~days) and planet sizes ($R_P=1-2 R_{Earth}$) with unprecedented fine detail.
We could precisely identify three climate state transition regions that are associated with phase transitions in standing tropical and extra tropical Rossby waves.

We confirm that the climate on fast rotating planets may assume multiple states ($P_{rot}\leq 12$~days for $R_P=2 R_{Earth}$). Our study is, however, the first to identify the type of planetary wave associated with different climate states: The first state is dominated by standing tropical Rossby waves with fast equatorial superrotation. The second state is dominated by standing extra tropical Rossby waves with high latitude westerly jets with slower wind speeds. For very fast rotations ($P_{rot}\leq 5$~days for $R_P=2 R_{Earth}$), we find another climate state transition, where the standing tropical and extra tropical Rossby wave can both fit on the planet. Thus, a third state with a mixture of the two planetary waves becomes possible that exhibits three jets.

Different climate states may be observable, because the upper atmosphere's hot spot is eastward shifted with respect to the substellar point in the first state, westward shifted in the second state and the third state shows a longitudinal 'smearing' of the spot across the substellar point.

We show, furthermore, that the largest fast rotating planet in our study exhibits atmosphere features known from hot Jupiters like fast equatorial superrotation and a temperature chevron in the upper atmosphere.
\end{abstract}


%\begin{document}
% Start your text

% Start your text
\section{Introduction}
\label{Introduction}
Transiting habitable Super-Earths around small, cool M~dwarf stars are easier to detect than other habitable planet-star configurations of the same stellar brightness due to their favourable star-planet contrast ratio (e.g. \cite{Madhusudhan2015}). They are thus the prime targets for the characterization of the atmospheres of terrestrial habitable exoplanets in the near future. These planets allow, furthermore, to test planet formation, evolution and planet atmosphere properties under conditions, for which there is no counterpart in the Solar System. Several upcoming space missions, such as TESS \citep{Ricker2014}, CHEOPS \citep{Broeg2013} and PLATO \citep{Rauer2014}, and ground based surveys (\citet{Snellen2012} and \citet{Gillon2013}, among others) are expected to yield candidates for the characterization of Super-Earths planets around bright nearby stars.

At least the planets at the inner edge of the habitable zone of M dwarfs are so close to their star that tidal interactions can force the planet's rotation in short astronomical time scales to be synchronous with the orbital period (e.g. \cite{Kasting1993}, \cite{Murray_Dermott}). They should thus continuously face their star with the same side; resulting in an eternal day and night side. Such a configuration is definitely not Earth-like and challenges our concept of 'habitability'. Consequently, there are still many uncertainties about the exact shape of planet atmosphere dynamics that such alien planets can assume. Parameter studies are one tool to investigate different climate dynamic regimes on tidally locked planets, but are very costly in computation time.

\cite{Joshi1997} were the first to investigate 3D large scale planet atmosphere dynamics on tidally locked habitable planets for selected parameters. They found that heat can be efficiently transported from the day side towards the night side for sufficiently dense atmospheres (surface pressure $p_S \geqslant 100$~mbar). They focused, however, on small Earth-size planets and did not change the rotation period.

Several more parameter studies have been performed in the last years that investigated various aspects of 3D planet atmosphere dynamics on tidally locked habitable planets with an Earth-like atmosphere: \citet{Edson2011} revealed an abrupt phase change in atmosphere dynamics on an Earth-size planet for rotation periods between $3-5$~days. \citet{HengVogt2011} investigated surface temperatures and wind structures for different surface friction, radiative time scale assumptions, surface gravity, surface pressures and rotation periods on a Super-Earth planet ($R_P=1.45 R_{Earth}$). They showed that different surface friction assumptions lead to very different surface wind and temperature structures. \cite{Merlis2010} investigated the general climate state for a tidally locked Earth-size ocean planet for very fast ($P_{rot}=1$~day) and very slow rotation ($P_{rot}=365$~days) and found that the atmosphere dynamics of the fast rotator is dominated by westerly jets at high latitudes, whereas the dynamics on the slow rotator is dominated by direct circulation from the substellar point to the night side. \cite{Carone2014} investigated the dynamics of a Super-Earth planet ($R_P=1.45 R_{Earth}$) for two rotation periods $P_{rot}=10$ and $P_{rot}=36.5$~days and showed, among other things, the existence of an embedded reverse circulation cell inside in a larger direction circulation cell for the faster rotation.

These investigations used models of different complexities and water content, focused on different aspect of atmosphere dynamics, and - what is more important - covered only a small part of the relevant planet rotation and planet size parameter space for tidally locked habitable planets around M dwarfs. The current state of knowledge makes it thus difficult to draw conclusions about the evolution of climate dynamics with increasing rotation. Even worse, climate state changes can be overlooked in a 'coarse' study, if they appear unexpectedly in a small part of the parameter space. It is, furthermore, difficult to identify the underlying mechanisms for climate state transitions as the usage of different planet parameters and model prescriptions act as confounding factors.

The purpose of this paper is, thus, to remedy this shortcoming: We will map the climate dynamic regime for terrestrial habitable planets around M dwarfs with hitherto unprecedented thoroughness, covering $P_{rot}=1 -100$~days and $R_P=1-2 R_{Earth}$. We will demonstrate that the perturbation method and the meridional Rossby wave numbers constitute very powerful diagnostic tools for the discussion of the simulation results. The two methods facilitate the precise identification of climate phase changes and multiple climate solutions for the same set of parameters. They allow us, furthermore, to identify the specific Rossby wave responsible for a given climate state. We can even coherently compare climate dynamics arising from different models and explain the likely origin of discrepancies.

In the following, we will first briefly describe the model and numerical adjustments. We will then introduce useful quantities like tropical and extra tropical Rossby radius of deformation, dynamical and radiative time scales, and the perturbation method. We will use these quantities to identify climate regime transitions due to different standing Rossby waves in the investigated parameter space. 

With an improved understanding of the Rossby wave climate regimes, we will then discuss the evolution of different atmosphere features like zonal wind jets, cyclonic vortices, the state of circulation cells, day side and night side surface temperatures, and vertical temperature structure. We will identify characteristic features that are associated with different standing Rossby wave regimes: the location of the upper atmosphere hot spot, the number and strength of zonal wind jets, and the state of circulation cells. 

We will then briefly discuss the potential to observationally discriminate between different climate dynamic states, provide a summary of our results and finish with a conclusion and outlook.

\section{The model}

Our model is described in details in \cite{Carone2014}. It is suitable for the 3D modelling of tidally locked terrestrial planets with zero obliquity. It can be used for different atmosphere compositions, planetary and stellar parameters (stellar effective temperature and stellar radius). The model was developed to be as versatile and computationally efficient as possible, using simple principles and the atmosphere properties of Solar System planets as a guide line. We will give here a brief prescription and explain the specific set-up used for this particular study.

\subsection{The dynamical core: MITgcm}

We employ the dynamical core of the Massachusetts Institute of Technology
global circulation model (MITgcm) developed at MIT\footnote{http://mitgcm.org} \citep{Adcroft2004}. It uses the finite-volume method to solve the primitive hydrostatical equations (HPE) can be written as the horizontal momentum equation
\begin{equation}
\frac{D\vec{v}}{Dt} +f\vec{k} \times \vec{v}=-\nabla_p \Phi +\mathcal{\vec{F}}_v,\label{eq: mom}
\end{equation}
vertical stratification equation
\begin{equation}
\frac{\partial \Phi}{\partial p}=-\frac{1}{\rho},
\end{equation}
continuity of mass
\begin{equation}
\vec{\nabla}_p\cdot \vec{v}+\frac{\partial \omega}{\partial p}=0,
\end{equation}
equation of state for an ideal gas
\begin{equation}
p=\rho \left(R/\mu\right)T,
\end{equation}
and thermal forcing equation
\begin{equation}
\frac{D\theta}{Dt}=\frac{\theta}{c_pT}\mathcal{F}_\theta,\label{eq:thermo}
\end{equation}
where the total derivative is
\begin{equation}
\frac{D}{Dt}=\frac{\partial}{\partial t}+\vec{v}\cdot \nabla_p+\omega \frac{\partial}{\partial p}\label{eq: thermal_forcing}.
\end{equation}
Here, $\vec{v}$ is the horizontal velocity and $\omega=Dp/Dt$ is the vertical velocity component in an isobaric coordinate system and, $p$ and $t$ are pressure and time, respectively. $\Phi=gz$ is the geopotential, with $z$ being the height of the pressure surface (also called geopotential height), and $g$ the surface gravity. $f=2\Omega \sin \nu$ is the Coriolis parameter, where $\Omega$ is the planet's angular velocity and $\nu$ the latitude at a given location. $\vec{k}$ is the local vertical unit vector, $T$, $\rho$ and $c_p$ are the temperature, density, and specific heat at constant pressure, respectively. $\theta=T\left(p/p_s\right)^\kappa$ is the potential temperature with $\kappa$ being the ratio of the specific gas constant $R/\mu$ to $c_p$. Here, $R=8.314$~JK$^{-1}$mol$^{-1}$ is the ideal gas constant, $\mu$ is the molecular mean mass of the atmosphere, and $p_s$ is the mean constant surface pressure defined at the surface. The pressure at surface level $p(0 mbar)$ varies horizontally around $p_s$ in the model. $\mathcal{F}_{\theta,v}$ are the horizontal momentum and thermal forcing terms, respectively. Note that $\mathcal{\vec{F}}_v$ is a vector and $\mathcal{F}_\theta$ is a scalar.

\subsection{Temperature forcing, surface friction and sponge layer}
\label{sec: forcing_friction}
The temperature forcing $\mathcal{F}_\theta$ is realized by using Newtonian cooling that drives the atmosphere temperature $T$ with a relaxation timescale $\tau_{rad}$ towards a prescribed equilibrium temperature $T_{eq}$, which writes in the easier recognizable temperature form as:
\begin{equation}
\mathcal{F}_T = \frac{T_{eq} -T}{\tau_{rad}}.
\end{equation}
$\tau_{rad}$ is calculated using basic estimates (e.g. \cite{Showman2002}), see also Section~\ref{sec: time scale}). The equilibrium temperature $T_{eq}$ is described in details in \cite{Carone2014}, where we have developed $T_{eq}$ for the day and night side. The illuminated dayside is driven towards radiative-convective temperature profiles suitable for a greenhouse atmosphere. The nightside temperature relaxes towards the Clausius-Clayperon relation of the main constituent of the atmosphere. Furthermore, we capped the top of the atmosphere with a vertically isothermal stratosphere.

The horizontal momentum forcing $\mathcal{\vec{F}}_v=\mathcal{\vec{F}}_{fric}+\mathcal{\vec{F}}_{sponge}$ consists of a surface friction and a sponge layer term:
\begin{equation}
\mathcal{\vec{F}}_{fric}= -\frac{1}{\tau_{fric}}\vec{v}
\end{equation}
and
\begin{equation}
\mathcal{\vec{F}}_{sponge}= -k_R \vec{v},
\end{equation}
where we use the Rayleigh friction approximation in both cases. The first term acts at the surface in the planetary boundary layer (PBL) between $p_s$ and $0.7\times p_s$. The surface friction time scale $\tau_{s,fric}$ is set to 1~day and linearly decreases with pressure until the top of the PBL is reached. The choices in boundary layer extent and surface friction time scale are taken over from \cite{HeldSuarez1994} for the Earth. However, we already discussed in \cite{Carone2014} that the prescription of the boundary layer varies greatly between models, even for the best studied other terrestrial planet: Mars. We will investigate changes in boundary layer extent and their effect on climate regime for tidally locked terrestrial paper in an upcoming paper. For now, we keep the Earth values from \cite{HeldSuarez1994} as described here.

The sponge layer term stabilizes the upper atmosphere boundary layer numerically against non-physical wave reflection. It may also be justified physically as a crude representation of gravity wave breaking at low atmosphere pressure. The sponge layer is implemented for $p\leq 100$~mbar, where $k_R$ increases from zero to $k_{max}$ at the upper atmosphere boundary using the prescription of \cite{Polvani2002} \citep{Carone2014}. In the nominal set-up $k_{max}$ was set to $1/80$~day${}^{-1}$. However, we found that the sponge layer efficiency needs to be increased in the fast rotation regime to ensure numerical stability. It will be shown later that this regime exhibits very fast equatorial superrotating jets with wind speeds larger than 100~m/s at the top of the atmosphere that suppresses upwelling over the substellar point. The dynamical interplay between these two dynamical features triggers more vigorous upper atmosphere dynamics that requires a stronger sponge layer to prevent un-physical wave reflection at the upper boundary of the model. The adjustment of sponge layer friction may tentatively be justified physically by assuming a more efficient wind braking mechanism for very fast wind speeds.  See Table~\ref{tab: tau_sponge} for the adopted $k_{max}$ dependent on rotation period. Conventionally, the sponge layer is omitted when discussing results because the layer is assumed to be not physically 'real'.

It should be noted, though, that $\mathcal{F}_T$ and $\mathcal{F}_{fric}$ are idealized representations of external heating and boundary layer physics, respectively.

\begin{table}
\caption{Sponge layer friction timescales used in our model}
\begin{tabular}{c|c|}
\hline
$k_{max}$ & $P_{rot}$ \\
{[}day${}^{-1}$] & [days]\\
\hline
1/80 & 4-100 \\
1/40 & 2-4 \\
1/20 & $\leq 2$\\
\hline
\end{tabular}
\label{tab: tau_sponge}
\end{table}

\subsection{Numerical setup}
We use the same set-up as described in \cite{Carone2014}. We use 20 vertical levels with equally spaced 50~mbar levels and surface pressure $p_s=1000$~mbar, to resolve the troposphere, that is, the weather layer of and Earth-like atmosphere as outlined in \cite{HeldSuarez1994}. Indeed, it was confirmed by \cite{Will1998} that this vertical resolution is sufficient to resolve the general structure of the Earth’s troposphere. This vertical resolution is consequently adopted by many climate studies of tidally locked Earth-like planets \citep{HengVogt2011,Edson2011,Merlis2010,Joshi1997,Joshi2003}. We use the C32 cubed-sphere grid for solving in the horizontal dimension \citep{Marshall2004}, which corresponds to a global resolution in longitude-latitude of $128\times 64$ or $2.8^{\circ}\times 2.8^{\circ}$. The substellar point is placed at 0 degrees longitude and latitude.

As established in \citet{Carone2014}, we evolve the atmosphere from initial conditions of universally constant temperature $T=264$~K for $t_{init}=400$~Earth~days. Data generated before $t_{init}$ are discarded; the simulation is run subsequently for $t_{run}=1000$~Earth~days and averaged over this time period. For $R_P\geq 1.25 R_{Earth}$, the nominal time step $\Delta t = 450$~s is applied, as used in \cite{Carone2014}. For the Earth-size planet simulations, $\Delta t = 300$~s is required to maintain the Courant - Friedrichs -Lewy (CFL) condition for the numerical stability of the finite volume method used in MITgcm. A reduction in planet size, while maintaining the C32 cube sphere grid, reduces the horizontal length scale and thus necessitates the reduction in time step.

\subsection{Investigated parameter space: Tidally locked habitable terrestrial planets around M dwarfs}
\label{sec: planet parameters}
We aim to investigate climate dynamics on tidally locked habitable planets around M dwarf stars more thoroughly than previous studies with respect to rotation period and planet size. At the same time, we strive to minimize confounding factors by keeping, e.g., thermal forcing constant across the investigated parameter space. The combination of a very thorough parameter sweep and the minimization of confounding factors allows to identify more climate dynamics phase changes and to better understand the underlying mechanisms behind them. It will be confirmed that in particular the rotation period is of great importance for the evolution of climate dynamics.

We keep for this study an Earth-like atmosphere composition: A nitrogen dominated greenhouse atmosphere with surface pressure $p_s=1000$~mbar. See Table~\ref{tab: Planetary parameters} for the relevant parameters \footnote{Note that we use $R_P=1.45 R_{Earth}$ instead of $R_P=1.5 R_{Earth}$ to be consistent with \cite{Carone2014}.}. We assume, furthermore, that each planet receives the same amount of net stellar energy than the Earth. Such a planet lies in any case in the habitable zone as investigated by, e.g., \citet{Kasting1993,Koppa2013,Yang2014,Zsom2013}.

We introduce the net incident stellar flux, because not all of the incident stellar energy is used to heat a planet's atmosphere; part of it is reflected back, e.g., by clouds and ice. In the grey two-stream greenhouse model used in this study to calculate radiative-convective equilibrium temperatures $T_{eq}$, we encapsulate the amount of reflected stellar energy in the planetary albedo $\alpha$. The net incoming stellar flux is thus
\begin{equation}
I_{net}=I_0(1-\alpha) \label{eq: I_net},
\end{equation}

where the incident stellar flux that the planet receives at semi major axis $a$ is:
\begin{equation}
I_0=\frac{L_*}{4\pi a^2}\label{eq: I_0},
\end{equation}
where $L_*$ is the stellar luminosity and $a$ can be calculated from the orbital period and stellar mass using Kepler's third law. For the Earth, the following values can be assumed $I_0=1368$~W/m${}^{-2}$ and $\alpha=0.3$, which yields $I_{net}=958$~W/m${}^{-2}$ \citep{Carone2014}.

This study is highly relevant in mapping the climate dynamics landscape in preparation for other GCMs that self-consistently calculate $\alpha$ by taking into account cloud and surface ice formation, but are too computationally costly to perform a detailed parameter sweep in orbital period. The usage of a fixed $I_{net}$ in the parameter study allows us - in contrast to other GCMs - to avoid changing $\alpha$, which would change thermal forcing. The elimination of changes in the thermal forcing as a confounding factor brings the dependence of climate dynamic regimes on rotation period to the fore. We can thus identify particularly interesting rotation periods that are 'turning points' for climate dynamics. We will explore in the following, for which orbital period range around M dwarfs it is indeed reasonable to assume $I_{net}=958$~W/m${}^{-2}$.

To derive luminosity and mass of M dwarfs, we use Yale-Yonsei isochrones for the low stellar mass regime with approximately solar values: an stellar age of 4 Gyrs, hydrogen mass fraction $X=0.70952$, $Z=0.01631$, metallicity $[Fe/H]=0.0$ and mixing length scaled over pressure scale height $\alpha_{MLT}=1.875$ \citep{Spada2013}. The Yale-Yonsei isochrones cover M dwarf spectral classes M0V-M6V with masses between $M_*=0.1 -0.6 M_{Sun}$. \cite{Kaltenegger2009} list smaller stars in their study but note that their masses are not well constrained.

\begin{figure}
\includegraphics[width=0.48\textwidth]{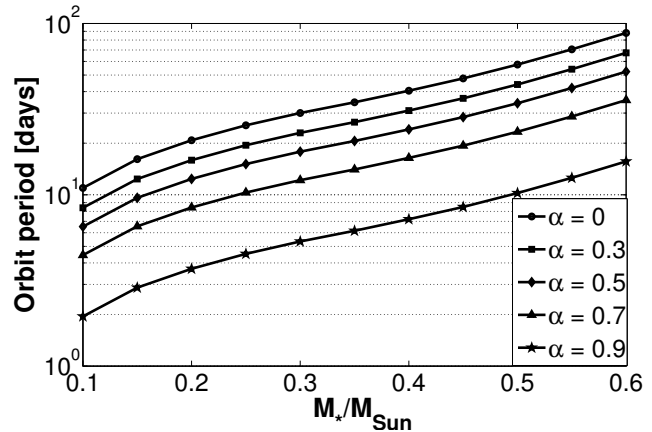}
\caption{Orbital period for tidally locked planets around M main sequence stars receiving $I_{net}=958$~W/m${}^{-2}$ for planetary albedo $\alpha=0,0.3,0.5,0.7,0.9$.}
\label{fig: Inet_Mdwarf}
\end{figure}

We can then calculate with Equations~(\ref{eq: I_net}) and (\ref{eq: I_0}), the orbital and thus rotation periods of tidally locked planets around M dwarfs that receive $I_{net}=958$~W/m${}^{-2}$, assuming $\alpha=0,0.3,0.5,0.7,0.9$. We find that Earth-like thermal forcing is indeed possible in the orbital period range $P_{rot}=1-100$~days for M dwarf stars\footnote{Older/younger M main sequence stars are more/less luminous than the stars investigated here. The corresponding planet orbits would thus shift towards/away from the star, respectively.}, if we allow extreme values of $\alpha$ (Figure~\ref{fig: Inet_Mdwarf}). \cite{Kaltenegger2009} derive $P_{rot}=2-65.5$~days by extending their M dwarf range to M9V stars, which is not so different from our rotation period range, but where the authors assumed a constant $\alpha=0.3$. While it is beyond the scope of this work to decide which planetary albedos are reasonable for tidally locked terrestrial planets around M dwarfs, we note that \cite{Leconte2013} consider $\alpha=0.1-0.5$, which corresponds to a relevant orbital period range between $P_{orb}=6-80$~days. Also, \cite{Yang2013} derive a planetary albedo as high as $\alpha=0.5$ for tidally locked planets around M dwarfs due to stabilizing cloud feedback above the substellar point, which extends the inner edge of the habitable zone towards the star.

Fast rotation periods with $P_{rot} \leq 6$~days allow us as a 'bonus' to study dynamical features already known from hot Jupiters. In particular, the largest planet size investigated in this study ($R_P=2 R_{Earth}$) will be of interest to bridge the gap between terrestrial planets and hot Neptunes. In other words, this study already lays the foundation for a coherent atmosphere dynamics study across the transition region between rocky and gas giant planets.

\begin{table}
\caption{Planetary and atmospheric parameters used in this study for all simulations}
\begin{tabular}{|l|c|}
\hline
Net incoming stellar flux $I_{net}$ & 958 W/m${}^2$\\
Obliquity & $0^\circ$\\
$c_p$ & 1.04 J/gK\\
$p_s$ & 1000 mbar \\
Main constituent & $N_2$\\
Molecular mass $\mu$ & 28 g/mol\\
Atmospheric emissivity $\epsilon$ & 0.8\\
Surface optical depth $\tau_s$ & 0.62 \\
Adiabatic index $\gamma$ & 7/5 \\
Surface friction timescale $\tau_{s,fric}$ & 1 day\\
Rotation period $P_{rot}$ & 1-100~days\\
\hline
\end{tabular}
\label{tab: Planetary parameters}
\end{table}

\begin{table}
\caption{Planetary size, surface gravity, and mass assuming Earth-like bulk density}
\begin{tabular}{c|c|c|}
\hline
$R_P$ & g & $M_P$\\
{[}$R_{Earth}$] & [m/s${}^2$] & [$M_{Earth}$]\\
\hline
1 & 9.8 & 1.0\\
1.25 & 12.5 & 2.0\\
1.45 & 14.3 & 3.1\\
1.75 & 17.3 & 5.4\\
2 & 19.6 & 8.0\\
\hline
\end{tabular}
\label{tab: planet_size}
\end{table}

The selected rotation period range $P_{rot}=1 - 100$~days was also investigated by \cite{Edson2011} and \cite{Merlis2010}. The authors focused, however, on an Earth-size planet, whereas we also investigate larger planets. \cite{Edson2011} scanned the whole rotation period region $P_{rot}=1 - 100$~days sparsely and focused on a specific rotation period region between $P_{rot}=3-5$~days, which the authors associated with climate bifurcation. That is, the authors inferred the existence of two different atmosphere dynamic states for the exact same set of planetary parameters and rotation periods. \cite{Merlis2010} investigated two rotation periods, $P_{rot}=1$~days and $P_{rot}=$~365~days. The latter case is dynamically very similar to the $P_{rot}=100$~days simulation of \cite{Edson2011}, the dynamics of which is dominated by direct circulation flow. We will show later that direct circulation is expected to dominate climate dynamics for slow rotators ($P_{rot}>>45$~days).

Recently, \cite{Leconte2015} questioned the common assumption that planets in the habitable zone of an M dwarf are tidally locked. However, the climate of tidally locked planets is very different from asynchronous rotating planets like the Earth. It gives rise to new interesting climate dynamics states, as we will show in this study. Furthermore, planets at the inner edge of the habitable zone are still probably tidally locked - even in the light of the revised tidal theory of \cite{Leconte2015}. Incidentally, these planets should be the first that become accessible to atmosphere characterization.

We investigate in this study planets of terrestrial composition with planet sizes between $R_P=1 -2 R_{Earth}$. This planet size range covers the Super-Earth planets, for which rocky composition can be safely assumed (e.g. \cite{Madhusudhan2012}). In addition, we assume Earth-like bulk density, $\rho_{Earth}=5.5$~g/cm${}^3$, for easier comparison between radiative and dynamical time scales and thus easier analysis of atmosphere dynamics evolution with planet size. Surface gravity can then be calculated via
\begin{equation}
g=\frac{G M_P}{R^2_P},
\end{equation}
where $G$ is the gravitational constant and planetary mass $M_P$ can be derived by:
\begin{equation}
M_P=\rho_{Earth} \frac{4}{3} \pi R_P^3,
\end{equation}
which yields planetary masses between 1-8 Earth masses (Table~\ref{tab: planet_size}).

We are aware that we probably underestimate the mass and thus the surface gravity of large rocky terrestrial Super Earth planets (see e.g. \cite{Hatzes2013}). We emphasize again, however, that we strive in this work to identify climate dynamics tendencies with rotation period and planet size. Thus, it is justified to minimize other confounding factors like differences in mean density between large and small terrestrial planets.  Furthermore, there are still many uncertainties in the composition of Super-Earth planets (e.g. \cite{Madhusudhan2012}, \cite{Rogers2010}).

\section{Atmosphere dynamics time and length scales}
\label{sec: time scale}
 For the discussion of climate state evolution with respect to rotation period and evolution, we now introduce basic atmosphere dynamics time and length scales. The most important time scales in the investigation of irradiated planet atmospheres are the dynamical $\tau_{dyn}$ and the radiative time scale $\tau_{rad}$. The former is  for tidally locked planets \citep{Showmanbook2011}
\begin{equation}
\tau_{dyn} \propto \frac{R_p}{U}\label{eq:tau_dyn},
\end{equation}
 where $U$ is the mean horizontal velocity or the horizontal velocity scale and the length scale $L$ is set to $R_P$, because circumglobal flow driven by the longitudinal heat gradient between the substellar point and the night side is assumed. The radiative time scale, as used in \cite{Carone2014}, is defined as
 \begin{equation}
\tau_{rad}= \frac{c_p p_s}{4 g \sigma T_s^3}\label{eq:tau_rad},
\end{equation}
where $T_s$ is the local surface temperature and $\sigma$ is the Stefan-Boltzmann constant. We discussed in \cite{Carone2014} the usefulness and limitation of this simple estimate and concluded that it is suitable for the modelling of the shallow atmosphere of terrestrial planets.

For the investigation of atmosphere dynamics with respect to planet rotation period, the most important parameter to monitor is the Rossby radius of deformation. Rossby waves are planetary waves that conserve absolute vorticity \citep{Holton} $\eta=\zeta +f$, where $\zeta=\vec{k} \cdot \vec{\nabla} \times \vec{v}=$ is the (vertical component of the) relative vorticity and $f=2 \Omega \sin \nu$ is the Coriolis parameter. Rossby waves owe their existence to variations of the Coriolis parameter with latitude and typically propagate in westward direction. The \textit{Rossby radius of deformation} represents the maximum amplitude of the planetary waves that can be excited at a given latitude.

The \textit{\textbf{extra tropical} Rossby radius} is defined as (e.g. \cite{Holton})
\begin{equation}
L_R=\frac{N H}{f} \label{eq: Rossby_mid},
\end{equation}
where $N=\sqrt{-\rho g^2/\theta \times \mathrm{d}\theta/\mathrm{d}p}$ is the buoyancy frequency in pressure coordinates that encapsulates the static stability of the atmosphere and $H=RT/\mu g$ is the scale height. We find the best agreement with changes in planet atmosphere dynamics, if we evaluate $L_R$ for $\nu=30^{\circ}$.

The \textit{\textbf{tropical} Rossby radius} is defined as
\begin{equation}
\lambda_R=\sqrt{\frac{NH}{2\beta}} \label{eq: Rossby_eq},
\end{equation}
where $\beta=2 \Omega/R_P $ in the equatorial beta plane approximation \citep{Holton}. We compare atmosphere dynamics for different ratios of the Rossby radius of deformation over the planet's radius $R_P$. It should be noted that we use here a different definition for Rossby radius of deformation than in \cite{Carone2014}, where we used the definition of \cite{Showman2011} for a shallow water model. It will be shown in this study that the definitions presented above, that are also used by \cite{Edson2011}, are more appropriate for a fully 3D~global circulation model.

We will show that the meridional Rossby wave numbers, that is, the ratios of the Rossby radii of deformation over the planet's radius $R_P$, $L_R/R_P$ and $\lambda_R/R_P$, are extremely important diagnostic tools in identifying possible climate state transitions with respect to rotation period and planet size. We will, thus, use these properties extensively to classify and discuss climate states.

\section{The perturbation method for Rossby wave analysis}
\label{sec: perturbation}
\begin{figure*}
\includegraphics[width=0.9\textwidth]{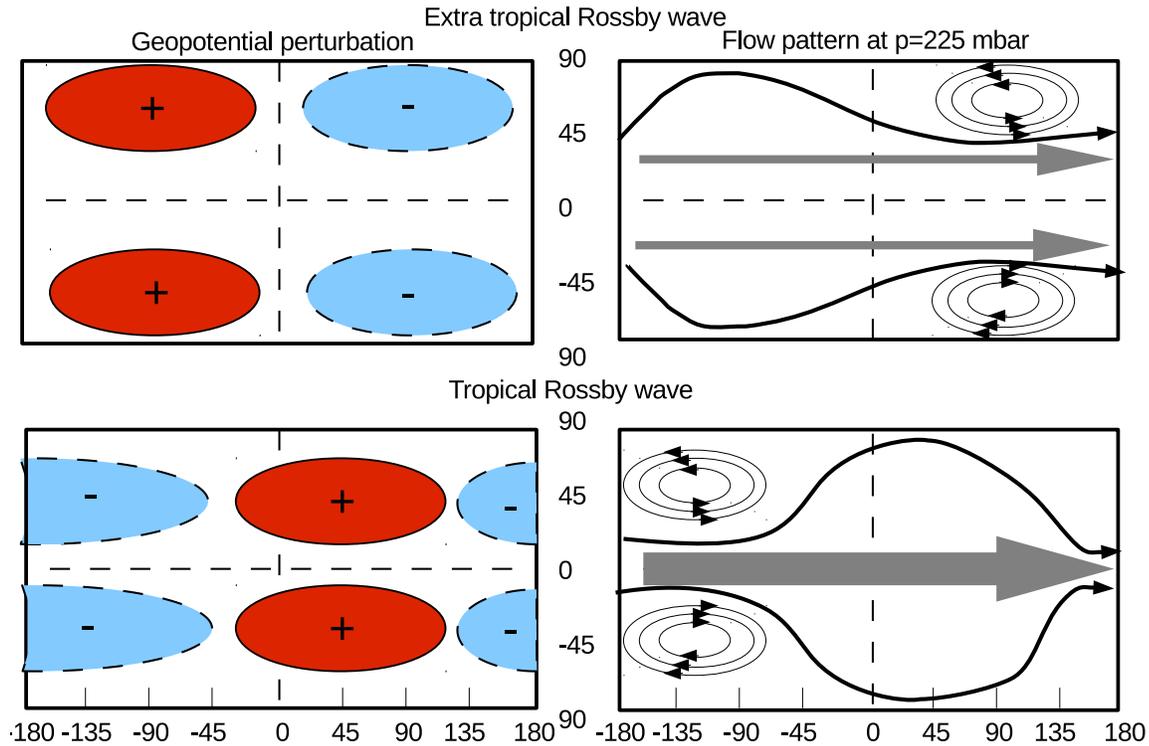}
\caption{Schematic idealized view of geopotential height perturbations and flow patterns at the top of the atmosphere ($p=225$~mbar) for the extra tropical (\textbf{top}) and extra tropical Rossby wave (\textbf{bottom}) and for meridional Rossby wave numbers close to unity, $L_R/R \approx 1$ and $\lambda_R/R \approx 1$, respectively. Red ellipses with black continuous circumferences denote positive gepotential height anomalies, light blue ellipses with dashed circumferences denote negative geopotential height anomalies. Grey arrows mark the strength, direction and approximate latitude position of zonal wind jets, the black arrows sketch the standing planetary Rossby waves with zonal wave number $n=1$ for each scenario. The thin black circles with arrow head mark the position of the cyclonic vortices associated with the negative geopotential height anomaly.}
\label{fig: Rossby_sketch}
\end{figure*}

For a better qualitative analysis of atmospheric waves, we follow the example of \cite{Edson2011} and \cite{Carone2014} and make use of the perturbation method \citep{Holton}. We will focus on perturbations in horizontal velocity $\vec{v}$ and geopotential height $z=\Phi/g$ as proxies for planetary waves. In perturbation theory, variables are divided into a basic state, which is assumed to be constant in time and longitude, and in an eddy part, which contains the deviations from the average.

We will focus in this study on temporally averaged flow and circulation to facilitate the interpretation of long-term climate states. We can therefore assume
\begin{equation}
\vec{v}=\bar{\vec{v}}+\vec{v}',
\end{equation}
and
\begin{equation}
z=\bar{z}+z',
\end{equation}
where the first term in each equation denotes the basic state, that is, the zonal mean of a given quantity, and the second term the eddy component, respectively. 

It will be shown that the perturbation method makes time averaged Rossby wave features directly visible and facilitates the interpretation of the physical processes that produce the observed large scale circulation patterns. The perturbation method in combination with the meridional Rossby wave numbers proves particularly powerful.

Since we focus in this study on contributions of standing tropical and extra tropical Rossby waves, we provide in the following a rough idealized sketch on expected geopotential height anomalies and resulting idealized flow patterns (Figure~\ref{fig: Rossby_sketch}), both for the respective meridional Rossby wave numbers close to unity.

Eddy velocities $\vec{v}'$ (not shown) exhibit cyclonic movement around negative and anti-cyclonic movement around positive geopotential height anomalies (Figure~\ref{fig: Rossby_sketch}, left panels). Eddy wind patterns circulating around one eddy geopotential height minimum and maximum per hemisphere are so called Rossby wave gyres for zonal wavenumber $n=1$. They are not centred at the equator - even when induced by the tropical Rossby wave \citep{Matsuno1966}. The extra tropical Rossby wave gyres, however, do tend to be shifted more polewards compared to their tropical counterpart (see e.g. \cite{Edson2011}). In cases with meridional Rossby wave numbers close to unity, which implies that the planetary wave engulfs the whole planet, it is, however, difficult to distinguish between both type of waves based on their latitudinal location. We found in this study that the longitudinal location of the geopotential height anomalies are a better indicator.

For tropical Rossby waves, we find Rossby wave gyres around negative geopotential height anomalies west from the substellar point and positive anomalies east from the substellar point (Figure~\ref{fig: Rossby_sketch}, bottom-left). Cyclonic vortices\footnote{'Cyclonic vortex' is a type of flow that has large positive relative vorticity $\zeta$ that is defined as the vertical component of the curl of the horizontal velocity $\zeta = \vec{k} \times \left( \vec{\nabla} \times \vec{v} \right)$\citep{Holton}. } in the observed flow patterns are thus located at longitude $-135^{\circ}$ over the negative geopotential height anomaly (Figure~\ref{fig: Rossby_sketch}, bottom-right). The longitudinal day and night side heat gradient is predicted to trigger a tropical Kelvin wave, which appears in the eddy velocity field as a purely zonal wind strip that is centred at the equator (not shown here). \cite{Showman2011} predicted that the shear between the tropical Kelvin and Rossby waves produces equatorial superrotation (Figure~\ref{fig: Rossby_sketch}, bottom-right).

In contrast to that, shear between the tropical Kelvin wave and extra tropical Rossby waves produces two westerly jets at mid-latitude that are weaker compared to the purely equatorial wave interaction scenario (Figure~\ref{fig: Rossby_sketch}, top-right). At the same time, the extra tropical Rossby wave gyres are shifted by almost 180 degrees with respect to the tropical Rossby wave gyres (Figure~\ref{fig: Rossby_sketch}, top-left). As a direct consequence, the vortices are now located at longitude $+90^{\circ}$ (Figure~\ref{fig: Rossby_sketch}, top-right) over the now eastward shifted eddy geopotential height minimum. To maximize the eddy geopotential height signal, we will apply the perturbation method to the upper atmosphere layer at $p=225$~mbar, following \citep{Edson2011,Carone2014} . We have verified that the sponger layer at $p=100$~mbar does not affect our results. We will also analyze the horizontal flow pattern at this altitude, which allows to directly tie changes in the Rossby waves to changes in the horizontal flow.

We find in this study either climate states dominated by tropical Rossby waves or a mixture between tropical and extra tropical Rossby waves. We will show in an upcoming paper, however, that it is indeed possible to achieve a climate state strongly dominated by extra tropical Rossby waves with our model, if surface friction is increased.

\section{Results and Discussion}
We will discuss the evolution of the horizontal flow from slow $P_{rot}=100$~days to very fast rotations $P_{rot}=1$~day and focus on the possible transition regions associated with phase changes in the Rossby waves excited in the planet's atmosphere. The latter are associated with changes in the meridional Rossby wave numbers, $L_R/R_P$ and $\lambda_R/R_P$. The transition in meridional Rossby wave numbers are inspected using the perturbation method. The combination of both methods allows to precisely identify which type of planetary wave is responsible for a given climate state.

We will show that we can identify more climate state transitions than previous studies due to our very thorough investigation of rotation period and planet size phase space. In particular, we identify two regions in the fast rotation regime that allow for the existence of multiple climate states. We will show that different climates states result in different zonal wind jet structures, longitudinal cyclonic vortex locations, and different the upper atmosphere hot spot locations with respect to the substellar point.

We will furthermore discuss circulation cell strength and number and their interaction with horizontal flow patterns in  different climate states. We will show that fast equatorial superrotation suppresses direct circulation and that large planets can exhibit fragmented circulation cells. We will also discuss surface temperature evolution in comparison with the previously obtained results, in particular, with the diminishing strength of the tropical direct circulation due to strong equatorial superrotation. Finally, we will investigate the vertical temperature and explain two different types of upper atmosphere temperature inversions based on the insights gathered from previous results.

\subsection{Zonal wind}
\begin{figure}
\includegraphics[width=0.48\textwidth]{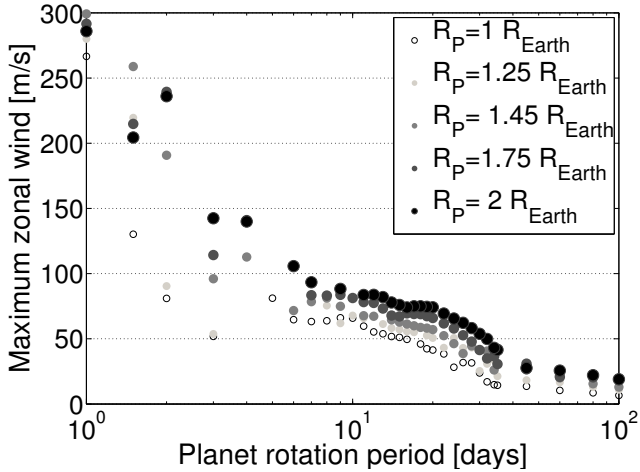}
\caption{Maximum zonal wind speeds in m/s on a tidally locked terrestrial planet for different planet radii ($R_P=1-2 R_{Earth}$) and rotation periods ($P_{rot}=1-100$~days), assuming identical thermal forcing, that is, moderate Earth-like irradiation.}
\label{fig: zonal_wind}
\end{figure}

Figure~\ref{fig: zonal_wind} shows the maximum zonal wind speeds with respect to planet rotation period and planet size. A few things are immediately apparent:
\begin{itemize}
\item Positive values indicate that the zonal wind maxima are associated with westerly winds, that is, the atmosphere flows in general in the direction of the planet's rotation.
\item Large planets tend to have larger wind speeds.
\item Wind speed increases for faster rotation periods starting with a few tens of m/s for the slowest rotation ($P_{rot}=100$~days) and reaching velocities larger than 100~m/s for very fast rotation ($P_{rot}=1$~day).
\item For fast rotating planets with $1.5 \leq P_{rot}\leq 5$~days, the wind speed maxima show a wide spread of possible values.
\item For very fast rotation, $P_{rot}=1$~day, the wind speeds appear to converge for all planet sizes to wind speeds between 275-300~m/s.
\end{itemize}

The westerly flow can be understood according to \cite{Showman2011} by the shear between equatorial (eastward travelling) Kelvin waves and tropical (westward) travelling Rossby waves that lead to net angular momentum transport from high to low latitudes and thus to westerly flow. The meridional tilt of the Rossby wave gyres, a clear indication of the Kelvin-Rossby wave coupling - will be shown later for fast rotating planets. The other zonal wind tendencies warrant a more detailed analysis.

\subsubsection{Dependency of zonal wind speeds on planet size}
\label{zonal_wind_Rp}

The tendency of larger planets to have larger wind speeds can be understood by performing a scale analysis on the thermodynamic energy equation. The first law of thermodynamics can be written for synoptic scales and horizontal temperature fluctuations \citep{Holton} as
\begin{equation}
\frac{T}{\theta_0}\left(\frac{\partial \theta}{\partial t}+u\frac{\partial \theta}{\partial x}+v\frac{\partial \theta}{\partial y}\right)=\frac{J} {c_P}\label{eq: thermodyn},
\end{equation}
if we neglect vertical transport of energy (see also Equation~\ref{eq: thermal_forcing}). $\theta_0(z)$ is the basic state of the potential temperature, $\theta(x,y,t)$ the deviation from it, where $z$ is the vertical component in the Cartesian coordinate system, and $J$ is the heating rate. Because we assume the same temperature forcing and atmosphere composition throughout simulations in this study, we can assume that $\partial \theta/\partial t$ is approximately constant for changes with planet size. Therefore, we can derive the planet size dependence of the remaining wind advection term on the left hand side of equation~(\ref{eq: thermodyn}) via scale analysis with \citep{Holton},
\begin{equation}
 \frac{\theta U}{L} \sim \frac{J} {c_P} \label{eq: scale analysis},
\end{equation}
 where $U$ is the horizontal velocity scale, and $L$ is the horizontal length scale. As already discussed in Section~\ref{sec: planet parameters}, $L=R_P$ (see also \cite{Showmanbook2011}).  $\theta(x,y)$ is in the context of tidally locked terrestrial planets the difference between day and night side temperature, which we prescribed as equal for all planet sizes. We use Newtonian cooling for the heating rate, so that
\begin{equation}
J= \frac{\Delta \theta}{\tau_{rad}}\label{eq: heating rate},
\end{equation}
where $\Delta \theta$ is the difference between the potential temperature $\theta$ in the atmosphere and the prescribed equilibrium potential temperature $\theta_{eq}$. $\tau_{rad}$ is again the radiative time scale. If we assume that $\Delta \theta$ hardly changes with planet sizes, then the heating rate $J$ dependency on planet size is determined via $\tau_{rad}$. The radiative time scale depends inversely on surface gravity, where $g\propto R_P$ from Section~\ref{sec: planet parameters}. From equations~(\ref{eq:tau_rad}) and (\ref{eq: heating rate}), it follows that $J \propto R_P$. Because also $L=R_P$, it follows ultimately from equation~(\ref{eq: scale analysis}) that wind speeds depend on planet size via relation:
\begin{equation}
U \propto R^2_P.
\end{equation}
Therefore, we expect stronger wind speeds by a factor of $4$ for the 2 Earth radii planet compared to the 1 Earth radii planet. This relation, however, only holds for the free atmosphere and if the Coriolis force is negligible, that is, for the atmosphere above the planet boundary layer and slow rotating planets (see, e.g. \cite{Showmanbook2011}).

\begin{table}
\caption{Zonal wind speeds in the troposphere and expected wind speeds $u_{exp}$ from scale analysis for $P_{rot}=100$~days and different planet sizes}
\begin{tabular}{c|c|c}
\hline
$R_P$ & $u_{max}$ & $u_{exp}$ \\
{[}$R_{Earth}$] & [m/s] & [m/s]\\
\hline
1 & 5 & 5 \\
1.25 & 11 & 8 \\
1.45 & 12 & 10.5\\
1.75 & 18.4 & 15.3\\
2 & 19.8  & 20\\
\hline
\end{tabular}
\label{tab: u_max_slow}
\end{table}

\begin{figure*}
\includegraphics[width=0.95\textwidth]{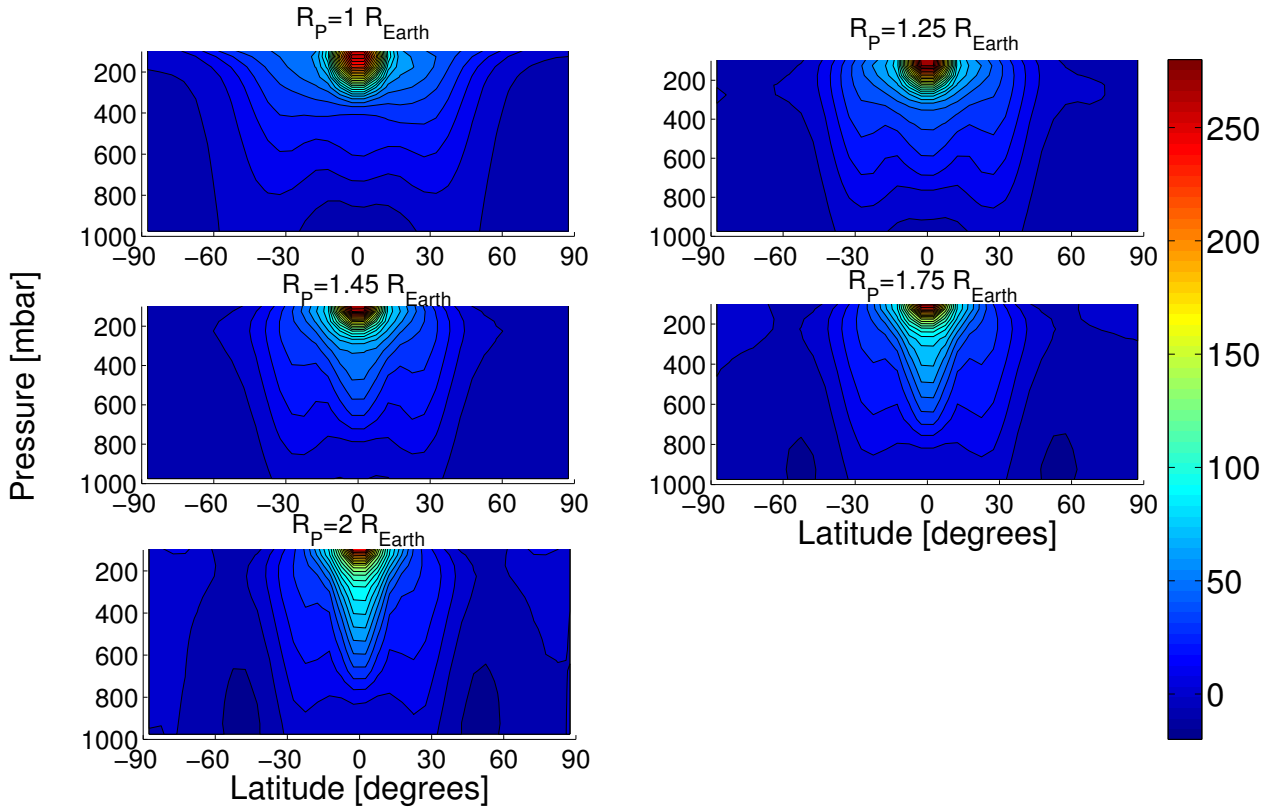}
\caption{Zonal wind speeds in m/s in our nominal climate model for $P_{rot}=P_{orb}=1$~day. Minimum and maximum wind speeds are -20 to 280 m/s, respectively. Contour interval is 10~m/s.}
\label{fig: zonal_Prot1d}
\end{figure*}

For $P_{rot}=100$~day, the maximum zonal wind speed in the troposphere of the $1 R_{Earth}$ planet is $u_{max}=5$~m/s and for the $2 R_{Earth}$ planet it is $u_{max}=19.8$~m/s. This is very close to the 1:4 relation expected from the scale analysis. The $R_P^2$ increase in zonal wind speeds is, indeed, maintained for the investigated planet sizes within a few m/s if $u_{max}=5$~m/s is taken as a basis value from the $R_P=1 R_{Earth}$ simulation (Table~\ref{tab: u_max_slow}). This is a good agreement between the scale analysis and the simulation results, if we assume that $u_{max} \propto U$. For faster rotation, however, pure advection from the hot day to the cold night side is no longer a valid assumption as the Coriolis force becomes a more dominant factor in shaping the flow. While bigger planets still tend to have higher wind speeds for faster rotation, the differences are much smaller than expected from $U\sim R^2_P$, and even appear to vanish for very fast rotation.

\subsubsection{Dependency of zonal wind speeds on planet rotation}

The increase of zonal wind speed with faster rotation is more complex and warrants an in-depth investigation. Generally, we see an increase of zonal wind speeds with faster rotation that reach maximum values of 300~m/s for $P_{rot}=1$~day. The high wind speeds of the very fast rotating planets can be directly attributed to equatorial super-rotation. Figure~\ref{fig: zonal_Prot1d} shows exemplary zonal mean of zonal winds for $P_{rot}=1$~day and different planet sizes.

Equatorial superrotation is a dynamical feature that frequently and consistently shows up in the atmosphere modelling of tidally locked hot Jupiters and Neptunes (e.g. \cite{Kataria2015}, \cite{Lewis2010}) and for hot Super-Earths (e.g. \cite{Kataria2014}, \cite{Zalucha2013}). Interestingly, atmosphere models of fast rotating tidally locked terrestrial exoplanets with habitable surface temperatures and the same nitrogen-dominated atmosphere composition are ambiguous with regard to equatorial superrotation. The simulations of \cite{Edson2011} also show zonal wind speed increase with faster rotation, but hardly exceed velocities of 100~m/s. Their simulations show two jets at high latitudes $\nu=\pm 60^{\circ}$ for fast rotation ($P_{rot}\leq 4$~days) instead of one equatorial jet. Even for $P_{rot}=1$~day, their simulations still have high latitude jets in addition to an equatorial superrotating jet with two maxima located at $\nu=\pm 10^{\circ}$. \cite{Merlis2010} performed a simulation for a tidally locked ocean planet with $P_{rot}=1$~day that exhibited two zonal jets at high latitude ($\nu=60^{\circ}$) and no equatorial superrotating jet. Both studies assumed an Earth-size planet. In contrast to that, all our simulations show efficient equatorial superrotation across all planet sizes.

\cite{Showman2011} attributed the emergence of equatorial superrotation and the associated fast wind speeds to the coupling between standing equatorial Kelvin and Rossby waves, provided the tropical Rossby radius of deformation $\lambda_R$ (Equation~\ref{eq: Rossby_eq}) is equal to or smaller than the planetary radius.

\begin{figure*}
\includegraphics[width=0.45\textwidth]{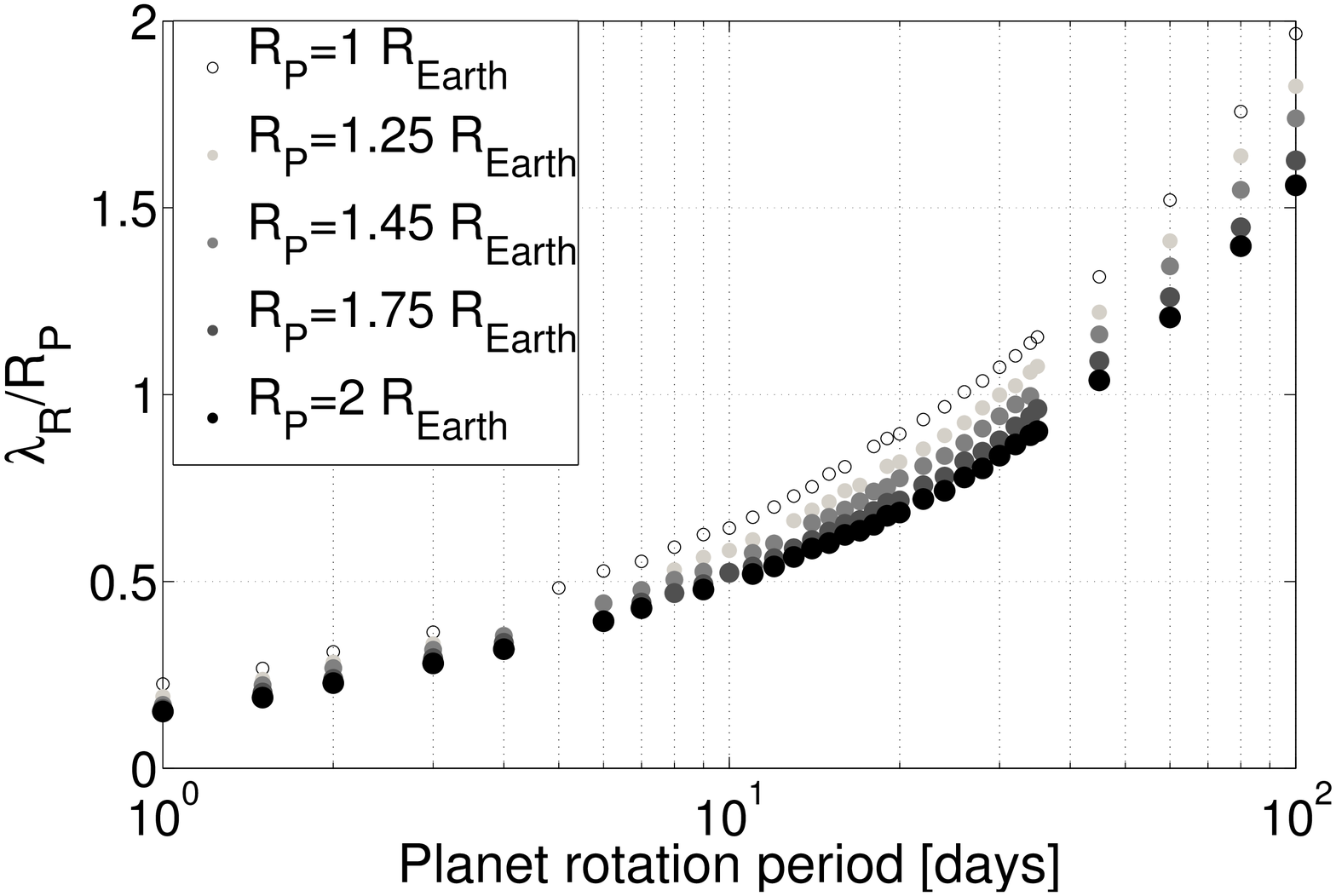}
\includegraphics[width=0.45\textwidth]{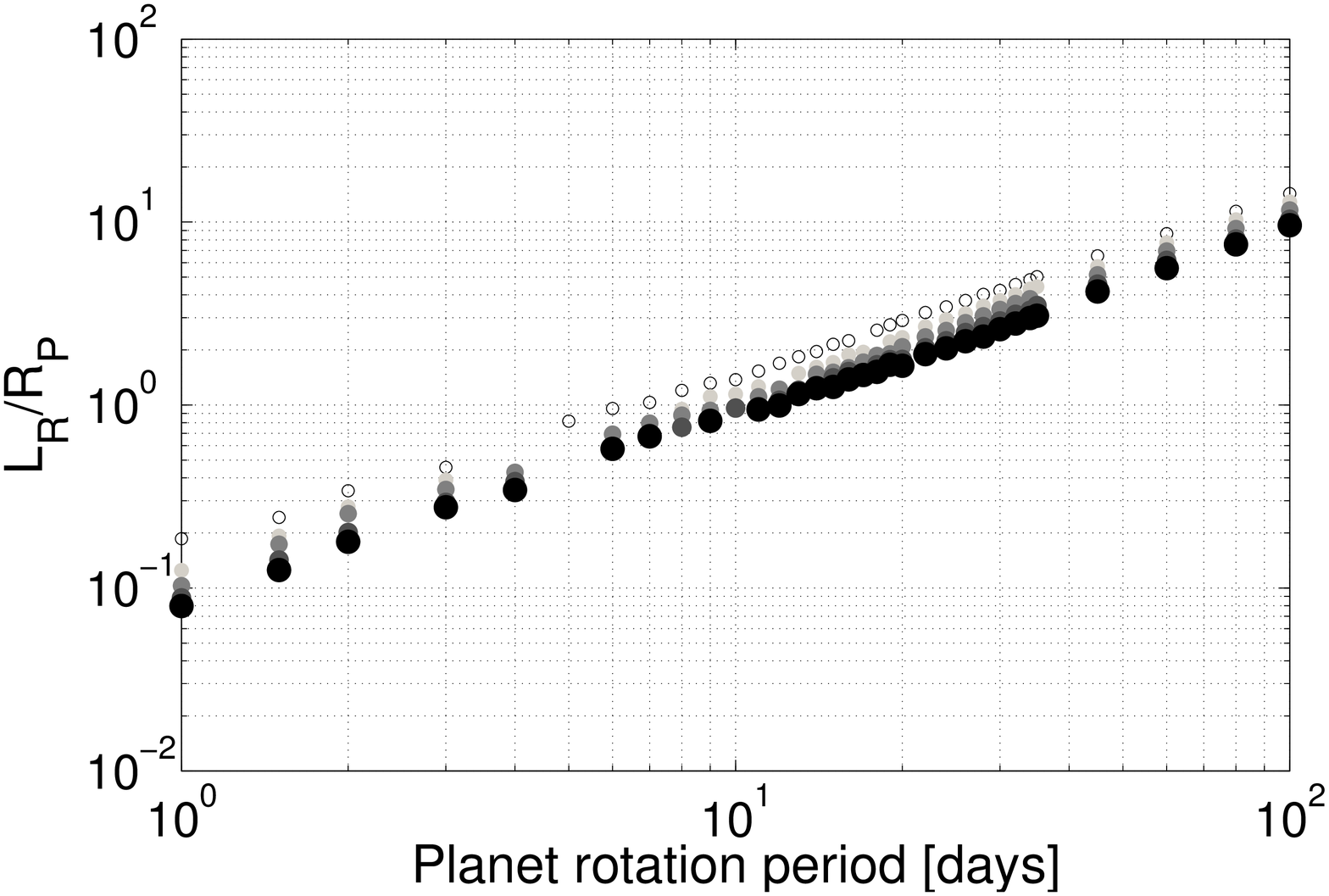}
\caption{\textbf{Left panel:} Equatorial Rossby radius of deformation $\lambda_R$ over planet radius $R_P$. \textbf{Right panel:} Mid-latitude Rossby radius of deformation evaluated at 30${}^\circ$ latitude over planet radius $R_P$. Both properties are shown for different planet sizes ($R_P=1-2 R_{Earth}$) and rotation periods ($P_{rot}=1-100$~days).}
\label{fig: Rossby_radius}
\end{figure*}

\cite{Edson2011}, on the other hand, speculated that the extra tropical Rossby radius of deformation $L_R$ (Equation~\ref{eq: Rossby_mid}) may be an important factor in shaping atmosphere dynamics in their simulations. They identified an abrupt transition in large scale circulation and zonal wind speeds at $P_{rot}=4.2$~days, which coincides with $L_R/R_P\approx 1$. Thus, the phase transition occurs according to \cite{Edson2011} because the extra tropical Rossby waves starts to 'fit' on the planet. It can be seen from their Table~2 that this phase transition coincides with another transition:  $\lambda_R/R_P\approx 0.5$. The latter may hint to an involvement of the equatorial Rossby wave as well. For rotations slower than the rotation period at the transition, that is, for $P_{rot}=5$~days, \cite{Edson2011} explicitly shows the presence of equatorial Rossby gyres in the eddy geopotential heights (their Figure~5g) and broad equatorial zonal wind jet feature (their Figure~4g). After transition ($P_{rot}=4$~days), the zonal wind jet and gyre structure abruptly changes to a high latitude configuration centred at $\nu=60^{\circ}$ (their Figures~4e and 5e).

In the light of the differences in wind jet structure for fast rotating planets from different climate models, we will study climate dynamics for different rotation periods in what follows. We will focus on standing Rossby waves and will show that we can link climate state transitions to either tropical or extra tropical Rossby wave transitions. The transitions can be inferred from the meridional Rossby wave numbers and further investigated with the perturbation method by means of the eddy wind and eddy geopotential structure at $p=225$~mbar. We will, furthermore, investigate how climate state transitions affect horizontal flow structure, where we also focus on the upper atmosphere at $p=225$~mbar.

\subsection{Phase transition in horizontal flow with meridional Rossby wave number}
\label{sec: phase transition}

As already pointed out by \cite{Edson2011}, it is not possible to exactly predict the behaviour of $L_R$ and $\lambda_R$ for a 3D planet atmosphere as both Rossby radii of deformation depend also on the vertical temperature structure via $d\theta/dp$ and on the surface temperatures via the scale height $H$. Both dynamical properties change as the atmosphere dynamics changes, as will be shown later. The meridional Rossby wave numbers $L_R/R_P$ and $\lambda_R/R_P$ not only decrease with faster rotation, they also depend on planet size (Figure~\ref{fig: Rossby_radius} and Table~\ref{tab: Rossby wave}).

Differences in surface temperature and vertical structure, as already pointed out in \citet{Carone2014}, explain why our $L_R/R_P\approx 1$ transition appears at slower rotation, $P_{rot}=6-7$~days, for $R_P=1 R_{Earth}$ compared to the model of \cite{Edson2011} that shows transition at $P_{rot}=4.2$~days (Figure~\ref{fig: Rossby_radius} and Table~\ref{tab: Rossby wave}).

\begin{table}
\caption{Rossby wave number transition for different planet sizes in our nominal model.}
\begin{tabular}{c|p{1.2cm}|p{1.2cm}|p{1.2cm}|p{1.2cm}}
\hline
$R_P$ & $P_{rot}$ for $\frac{L_R}{R_P}\approx 0.5$  & $P_{rot}$ for $\frac{L_R}{R_P}\approx 1$ & $P_{rot}$ for $\frac{\lambda_R}{R_P}\approx 1$& $P_{rot}$ for $\frac{\lambda_R}{R_P}\approx 0.5$\\
{[}$R_{Earth}$] & [days] & [days] & [days] & [days]\\
\hline
1 & 3-4& 6-7 & 25 & 5-6\\
1.25 & 4 & 8-9 & 30  & 7\\
1.45 & 5 & 10 &34 & 8\\
1.75 & 5-6& 11-12 &35-40 &9\\
2 & 5-6 & 12& 40-45 & 10-11\\
\hline
\end{tabular}
\label{tab: Rossby wave}
\end{table}

In the following, we will examine horizontal flow patterns at four transition regions in meridional Rossby wave numbers (Table~\ref{tab: Rossby wave}), where we will show that the second and third transition almost coincide and can thus not be investigated separately:
\begin{itemize}
\item $\lambda_R/R_P\approx 1$
\item $L_R/R_P\approx 1$
\item $\lambda_R/R_P\approx 0.5$
\item $L_R/R_P\approx 0.5$
\end{itemize}

\subsubsection{Rossby wave transition region: $\lambda_R/R_P\approx 1$}
\label{sec: lambda_1}
The transition region where the meridional extent of the tropical Rossby wave starts to 'fit' on the planet can be found between 25~days for an Earth-size and 45~days for a planet twice the size of Earth (Table~\ref{tab: Rossby wave}). When comparing with Figure~\ref{fig: zonal_wind}, this region coincides also with an increase in zonal wind speed acceleration with shorter rotation periods: The $\lambda_R/R_P>1$ region with $P_{rot}=45-100$~days, is a region of low zonal wind speeds between 10-30~m/s. The $\lambda_R/R_P \leqslant 1$ region with $P_{rot}=20-35$~days, shows an increase in zonal wind speeds to velocities of up to 50-75~m/s, depending on the planet size. This acceleration is in line with the interpretation that the equatorial Rossby waves form standing waves and transport angular momentum more efficiently from high to low latitudes. Also \cite{Leconte2013} pointed out that their 3D simulations show direct circulation states for $\lambda_R/R_P >> 1$ .

\begin{figure*}
\includegraphics[width=0.9\textwidth]{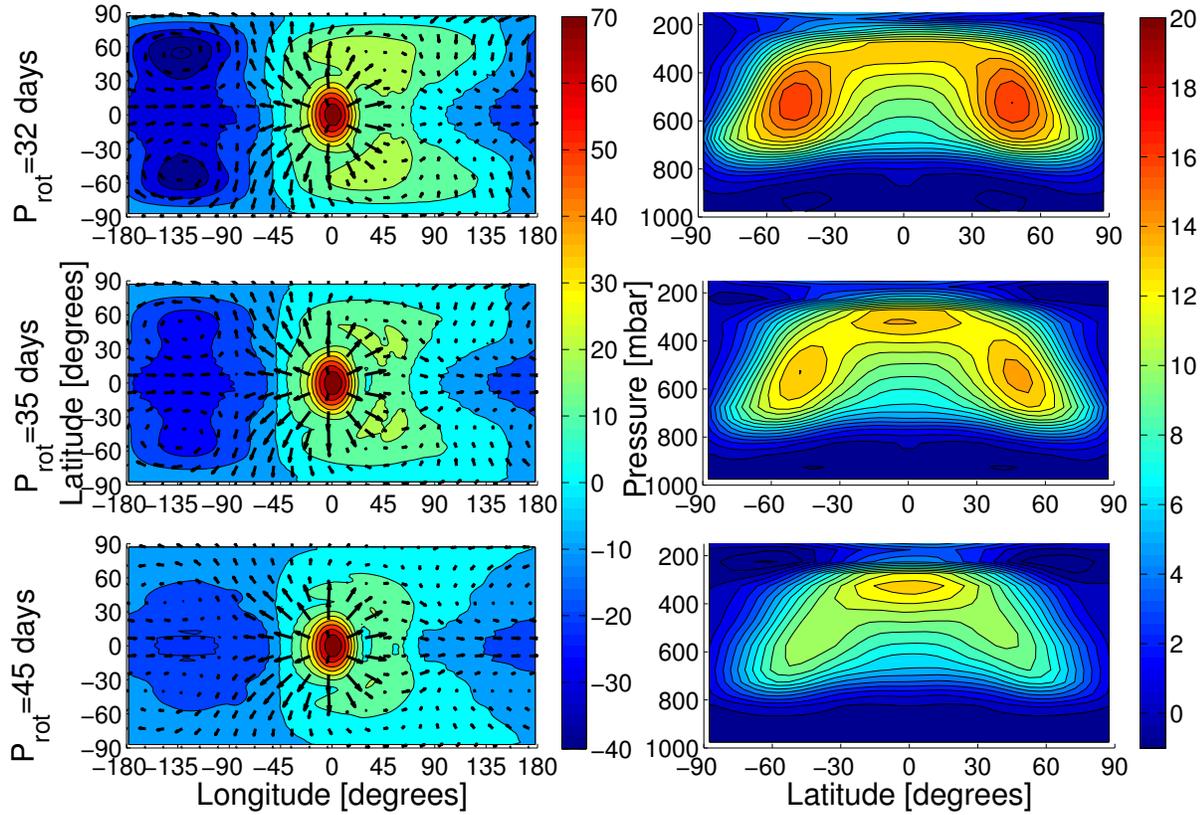}
\caption{\textbf{Left panel}: Eddy geopotential height in [m] and eddy horizontal winds in [m/s] at p = 225 mbar. \textbf{Right panel}: Zonal mean of zonal winds in [m/s]. Both are shown for an Earth-size planet ($R_P=1 R_{Earth}$) and rotation periods at the Rossby wave transition region  $\lambda_R/R_P\approx 1$ ($P_{rot}=32,35$, and 45~days for top to bottom rows). The longest wind vectors are from top to bottom: 27.6, 24.3, and 32.2 m/s, respectively. Contour intervals are 10~m for the eddy geopotential and 1 m/s for zonal wind speeds, respectively.}
\label{fig: lambda1_Rp1}
\end{figure*}

\begin{figure}
\includegraphics[width=0.45\textwidth]{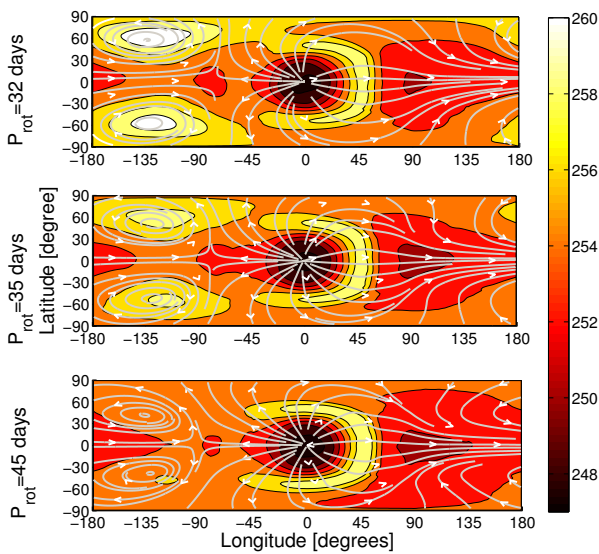}
\caption{Temperatures (contour interval 10 K) and streamlines of the horizontal flow, averaged over 1000 d, for pressure level p = 225 mbar for an Earth-size planet ($R_P=1 R_{Earth}$) and rotation periods at the Rossby wave transition region  $\lambda_R/R_P\approx 1$ (from top to bottom: $P_{rot}=32,35$, and 45~days).}
\label{fig: p225mb_lambda1_Rp1}
\end{figure}

Inspection of the eddy geopotential height $z'$ and eddy horizontal winds $\vec{v}'$ (see Section~\ref{sec: perturbation}) at the top of the atmosphere $p=225$~mbar shows indeed in this planet rotation region a gradual change from a climate state dominated by direct circulation to one where a standing tropical Rossby wave appears and starts to contribute to angular momentum transport - hence the zonal wind speed increase (Figure~\ref{fig: lambda1_Rp1}, left panel). Direct circulation is identifiable by divergent flow and a maximum in $z'$ at the substellar point, which marks the top of the direct circulation cell with horizontal transport from the hot substellar point towards the cool night side aloft and upwelling flow at the hot substellar point below. The standing tropical Rossby wave is identifiable by the Rossby wave gyres with negative eddy geopotential height east of the substellar point and positive west of it (See Figure~\ref{fig: Rossby_sketch}). 

One of the climate states discussed in \cite{Carone2014}, the Super Earth planet with $R_P=1.45 R_{Earth}$, $g=14.3$~m/s${}^2$ and $P_{rot}=36.5$~days resides in the transition regime. It was likewise shown to exhibit a mixture between superrotating flow and divergent flow at the top of the troposphere. It was then tentatively assumed that there is a connection between cyclonic vortices, the standing tropical Rossby wave and the zonal wind structure. This connection can be confirmed in this study: Although, weak cyclonic vortices are present also for slow rotations (e.g. $P_{rot}$=45~days), the cyclonic vortices gain in strength and become locations associated with upper atmosphere temperature maxima at the same time as the planetary wave starts to form a standing wave (Figure~\ref{fig: lambda1_Rp1}, right panel and Figure~\ref{fig: p225mb_lambda1_Rp1} ). 

\begin{figure*}
\includegraphics[width=0.9\textwidth]{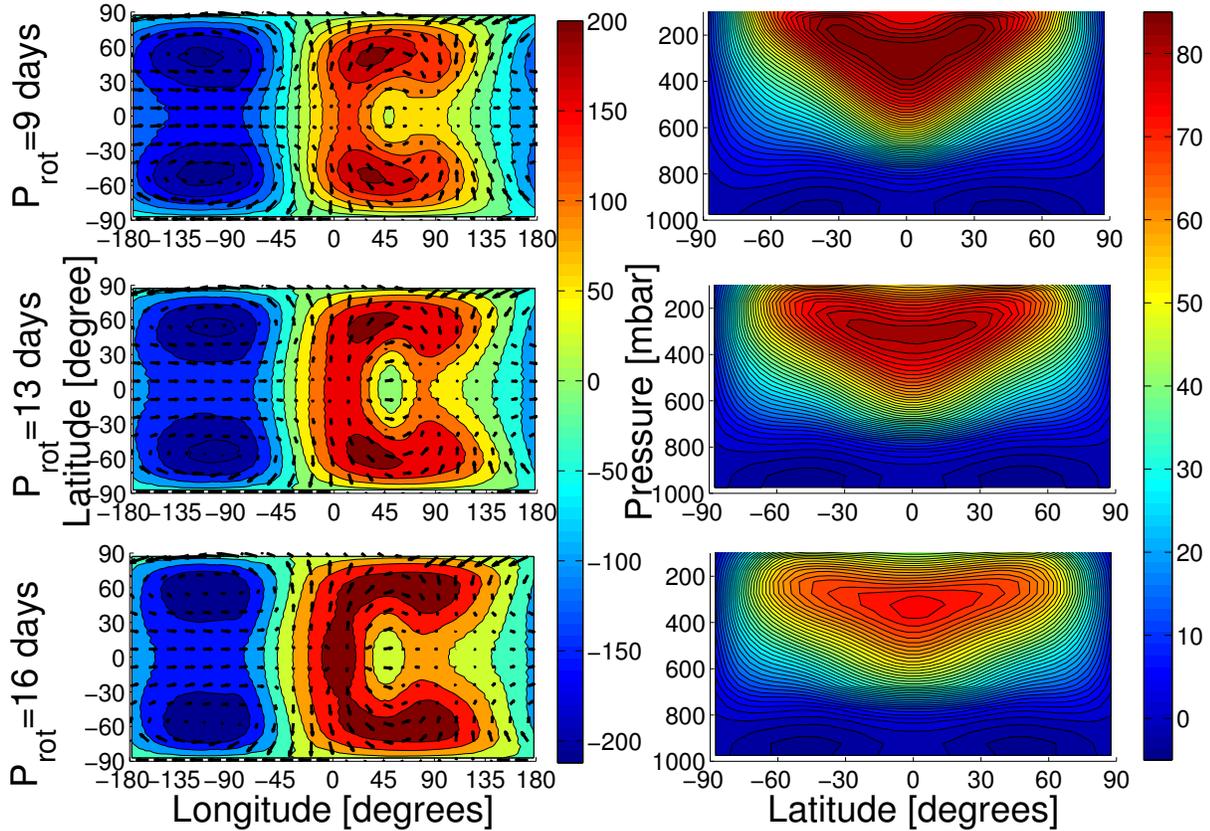}
\caption{Eddy geopotential height in [m] and eddy horizontal wind in [m/s] at p = 225 mbar (left panel) and zonal mean of zonal winds in [m/s] (right panel) for a large Super-Earth planet ($R_P=2 R_{Earth}$) and rotation periods at the Rossby wave transition region  $L_R/R_P\approx 1$ and $\lambda_R/R_P\approx 0.5$ ($P_{rot}=9,13$, and 16~days). The longest wind vectors are from top to bottom: 62.6, 59.0, and 61.1 m/s, respectively. Contour intervals are 50~m for the eddy geopotential and 2 m/s for zonal wind speeds, respectively.}
\label{fig: L1_Rp2}
\end{figure*}

In \cite{Carone2014}, we speculated that the temperature maxima associated with the vortices (Figure~\ref{fig: p225mb_lambda1_Rp1}) arouse from adiabatic heating of air masses streaming from the substellar point downwards into the vortices: The substellar point is a region of particularly high geopotential height. At the same time, the air emerging at the upper troposphere over this point is particularly cool, which can be explained by adiabatic cooling during the upwelling process. These cool air masses 'fall' downwards towards the vortices, which are cold features in the mid-atmosphere at $p=525$~mbar (see \cite{Carone2014}) and show up as negative anomalies in the geopotential height field. Consequently, the air is adiabatically heated over the vortices. We will show in Section~\ref{sec: vertical temperature} that the temperature maxima over the vortices cause temperature inversions in the vertical temperature structure, as already pointed out in \citet{Carone2014}. 

The transition to a standing tropical Rossby wave regime is also accompanied with a transition in wind structure: From a weak equatorial superrotation towards a broad zonal wind structure with two zonal wind maxima at mid-latitude and mid-troposphere ($p= 525$~mbar) (Figure~\ref{fig: lambda1_Rp1}, right panel). Apparently, angular momentum is indeed more efficiently transported from the poles towards lower latitudes with transition to $\lambda_R/R_P\leqslant 1$, as expected from shear between tropical Rossby waves and Kelvin waves. The latter are visible in the eddy velocity wind field as very confined equatorial stripe (between $\nu=\pm 10^{\circ}$) of purely zonal eddy winds, except over the substellar point where divergent upwelling flow superposes the Kelvin wave structure (Figure~\ref{fig: lambda1_Rp1}, left panel).

According to Figure~\ref{fig: lambda1_Rp1}, the emergence of the standing tropical Rossby wave can be found by visual inspection at $P_{rot}=35$~days instead of $P_{rot}=25$~days based on $\lambda_R/R_P$ (Table~\ref{tab: Rossby wave}). $P_{rot}=35$~days corresponds, however, to a $\lambda_R/R_P\approx 1.15$, which is only slightly larger than unity. Thus, we confirm that Rossby wave numbers give a good estimate but albeit not an exact guideline to identify changes in climate states. 

Comparison between the eddy geopotential height for $R_P=1 R_{Earth}$ and $P_{rot}=35$~days (Figure~\ref{fig: lambda1_Rp1}), with the eddy geopotential height for $R_P=1.45 R_{Earth}$ and $P_{rot}=36.5$~days (Figure~7, right panel in \cite{Carone2014}) confirms the planet size dependency. The Rossby wave gyres are more apparent for the Super-Earth-size planet than for the Earth-size planet, even though the rotation of the tidally locked Super-Earth discussed in \cite{Carone2014} is slightly slower. This is in line with the prediction that the rotation period for which standing equatorial Rossby waves emerge is shifted towards slower rotations with increasing planet size.

\subsubsection{Rossby wave transition region: $L_R/R_P\approx 1$ and $\lambda_R/R_P\approx 0.5$ }
\label{sec: L/R-1}

This is the regime that \cite{Edson2011} associated with multiple solutions for the climate state and with abrupt transition in the zonal wind speeds. Like in \cite{Edson2011}, the tropical and extra tropical Rossby wave transition appear to almost coincide for all planet sizes (Table~\ref{tab: Rossby wave}). Furthermore, another possible phase transition for the extra tropical Rossby wave with $L_R/R_P\approx 0.5$ is not very far away for an Earth size planet at $P_{rot}=3-4$~days. For the largest planet in our sample ($R_P=2 R_{Earth}$), the $L_R/R_P\approx 0.5$ transition lies at $P_{rot}=$5-6~days and thus farther away from the $L_R/R_P\approx 1$ transition at $P_{rot}=12$~days and the $\lambda_R/R_P\approx 0.5$ transition at $P_{rot}=10-11$~days, respectively. We will thus focus in the following discussion on atmosphere dynamics on the largest planet in our sample.

Inspection of zonal wind speed maxima (Figure~\ref{fig: zonal_wind}) shows that while the velocities still generally increase with faster rotation, the increase becomes less steep between $9 \leq P_{rot}\leq 20$~days. As soon as the equatorial Rossby wave transits to $\lambda_R/R_P < 0.5$ , the wind speeds rise again steeply. We do not find, however, an abrupt transition in wind speeds as reported by \cite{Edson2011}. This difference in zonal wind evolution will become clearer in the next section, as we explore why our model goes into full equatorial superrotation and the model of \cite{Edson2011} does not.

\begin{figure}
\includegraphics[width=0.45\textwidth]{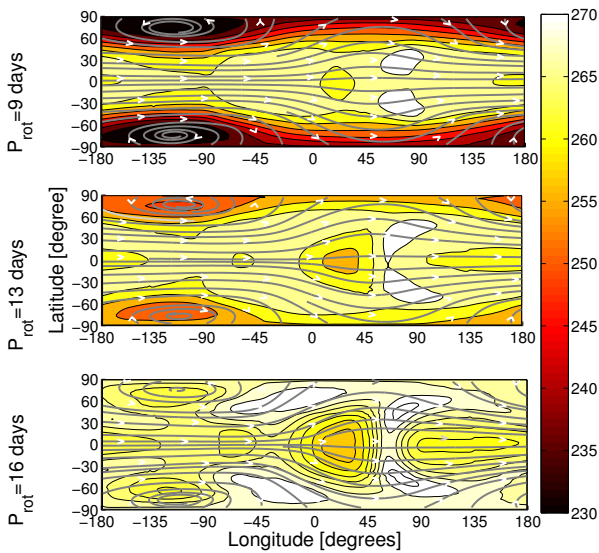}
\caption{Temperatures and streamlines of the horizontal flow, averaged over 1000 d, for pressure level p = 225 mbar for a large Super-Earth planet ($R_P=2 R_{Earth}$) and rotation periods at the Rossby wave transition region  $L_R/R_P\approx 1$ and $\lambda_R/R_P\approx 0.5$ (from top to bottom: $P_{rot}=9,13$, and 16~days). Contour intervals are 2~K for $P_{rot}=16$~days and 5~K for $P_{rot}=9$, and 13~days.}
\label{fig: p225mb_L1_Rp2}
\end{figure}

Figures~\ref{fig: L1_Rp2} and \ref{fig: p225mb_L1_Rp2} show indeed a change in climate state near the $\lambda_R/R_P\approx  0.5$ transition that affects the upper branch of the direct circulation cell. Upwelling from the substellar point can again be recognized as a cold spot at $p=225$~mbar\footnote{As noted in \cite{Carone2014}, the cool temperatures arise due to adiabatic cooling during uplifting.} (Figure~\ref{fig: p225mb_L1_Rp2}, bottom panel) and as a positive geopotential height anomaly with outstreaming flow in the eddy geopotential height and winds (Figure~\ref{fig: L1_Rp2}, left-bottom panel) for $P_{rot}=16$~days. After $\lambda_R/R_P\approx  0.5$, upwelling and divergence weaken and disappear as the Rossby wave gyres become even stronger (Figure~\ref{fig: p225mb_L1_Rp2}, top panel and Figure~\ref{fig: L1_Rp2}, left-top panel). At the same time, the zonal wind jet maxima converge towards the equator and increase in strength (Figure~\ref{fig: L1_Rp2}, right panel). It will be shown in Section~\ref{sec: circulation} that direct circulation and strong equatorial superrotation, indeed, don't mix very well: If direct circulation is strong, superrotation is diminished and vice versa.

With the disappearance of the divergence, the temperature of the cyclonic vortices at $p=225$~mbar - that have been pushed towards the poles with faster rotation - changes drastically. For $P_{rot}=16$~days, the vortices are still relatively warm, for $P_{rot}=9$~days, however, they are instead associated with very cool temperatures even at this low pressure levels. This result can again be explained in the context of divergent flow emerging from the substellar point 'falling down the slopes' of the warm - and thus vertically extended - equatorial superrotating region. Upwelling flow that falls into the vortices with low geopotential height at the top of the troposphere becomes, thus, adiabatically heated (see previous Section): provided the divergent flow can reach the increasingly pole-wards shifted vortices and provided there is divergent flow to heat up in the first place. The removal of divergent flow with faster rotation also explains why the horizontal temperature gradient increases between $P_{rot}=16$~days and $P_{rot}=10$~days at the top of the atmosphere (Figure~\ref{fig: p225mb_L1_Rp2} and also Figure~\ref{fig: p225mb_lambda1_Rp1} for comparison). We will take up the discussion of upper atmosphere thermal inversion again in Section~\ref{sec: vertical temperature}.

The atmosphere dynamics of our $P_{rot}=9$~days simulation for $R_P=2 R_{Earth}$ is, furthermore, very similar to the $P_{rot}=10$~days simulation in \citet{Carone2014} for $R_P=1.45 R_{Earth}$ and the $P_{rot} = 5$~days simulation in \cite{Edson2011}. All these cases lie near their respective $\lambda_R/R_P=0.5$ transition regime, which shows again how valuable the meridional Rossby wave numbers are as a diagnostic tool as they allow to compare results from different climate models with each other.

\begin{figure*}
\includegraphics[width=0.89\textwidth]{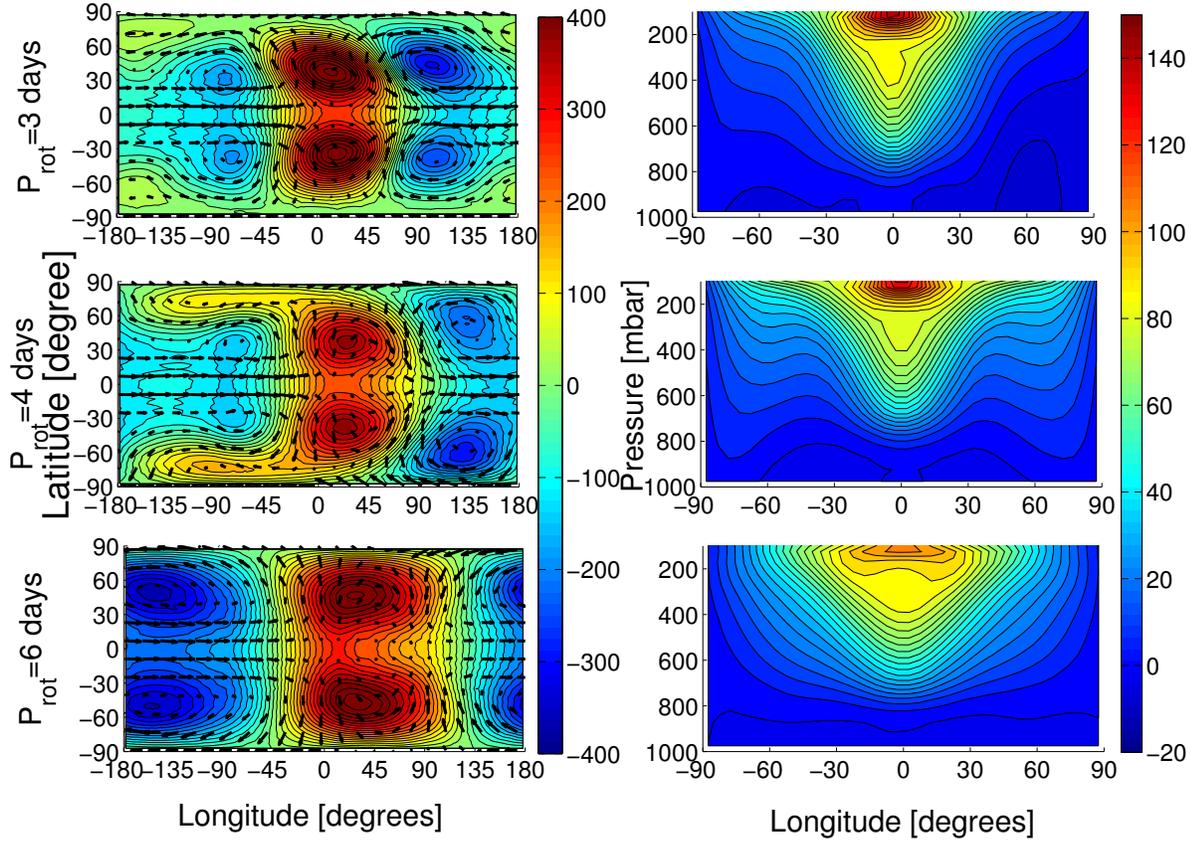}
\caption{Eddy geopotential height in [m] and eddy horizontal wind in [m/s] at p = 225 mbar (right panel) and zonal mean of zonal winds in [m/s] (left panel) for a large Super-Earth planet ($R_P=2 R_{Earth}$) and rotation periods at the Rossby wave transition region  $L_R/R_P\approx 0.5$ ($P_{rot}=3,4$, and 6~days). The longest wind vectors are from top to bottom: 61.5, 54.7, and 63.5 m/s, respectively. Contour intervals are 25~m for the eddy geopotential and 5 m/s for zonal wind speeds, respectively.}
\label{fig: L05_Rp2}
\end{figure*}

\begin{figure}
\includegraphics[width=0.45\textwidth]{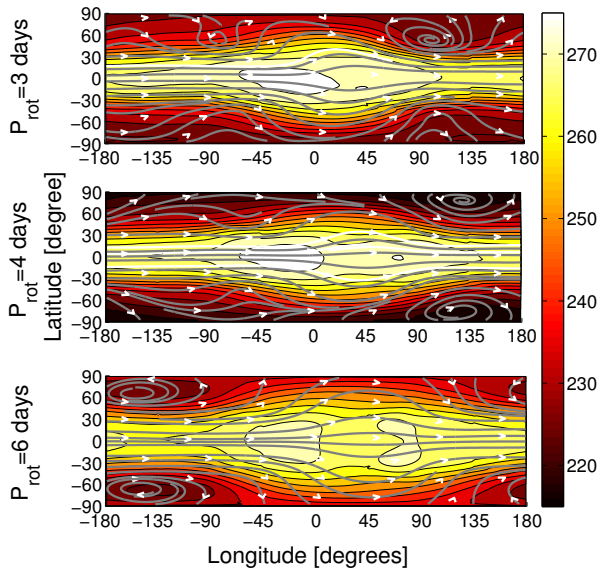}
\caption{Temperatures and streamlines of the horizontal flow, averaged over 1000 d, for pressure level p = 225 mbar, $R_P=2 R_{Earth}$ and rotation periods at the Rossby wave transition region  $L_R/R_P\approx 0.5$ (from top to bottom: $P_{rot}=3,4$, and 6~days). Contour intervals for temperatures are 5~K.}
\label{fig: p225mb_L05_Rp2}
\end{figure}

The increase in zonal wind speeds, the evolution of the zonal wind jet towards the equator and the deepening of the tropical Rossby wave gyres in the eddy wind and eddy geopotential height field show that the standing tropical Rossby waves are still the driving mechanism in this climate state regime. This is remarkable because the extra tropical meridional Rossby wave number ($L_R/R_P\leq 1$) shows that also the formation of standing extra tropical Rossby waves is, in principle, possible. The latter apparently don't yet play a role - at least in our nominal model. We will show in an upcoming paper, that an increase in surface friction triggers indeed the formation of standing extra tropical Rossby waves for $L_R/R_P\leq 1$.

\subsubsection{Rossby wave transition region: $L_R/R_P\approx 0.5$}
\label{sec: L/R-0.5}

The next climate state transition occurs for a large terrestrial planet with $R_P=2 R_{Earth}$ at $P_{rot}=5-6$~days. Indeed, the eddy geopotential height and eddy winds as well as the zonal wind structure show again a change in dynamics between $P_{rot}=6$~days and $P_{rot}=4$~days (Figures~\ref{fig: L05_Rp2} and \ref{fig: p225mb_L05_Rp2}).

Extratropical Rossby wave gyres appear besides the tropical Rossby wave gyres for $P_{rot}=4$~days. These can be identified by the apperance of additional geopotential height anomalies at high latitudes that are furthermore shifted in longitude by almost $180^{\circ}$ with respect to the tropical Rossby wave gyres (See also Figure~\ref{fig: Rossby_sketch}). That is, the extra tropical positive $z'$ structure is located west from the substellar point and the negative east from the substellar point, respectively (Figure~\ref{fig: L05_Rp2}, middle panel).

The extratropical standing Rossby waves lead to the formation of additional high latitude wind jets at $\nu=60^{\circ}$ (Figure~\ref{fig: L05_Rp2} and \ref{fig: p225mb_L05_Rp2}, middle panel). Because the equatorial and the extratropical Rossby wave have both meridional extent of less than half the planetary radius, \textbf{both} waves can now form standing planetary waves. The existence of two planetary waves can also be inferred from the horizontal flow streamlines at the upper troposphere (Figure~\ref{fig: p225mb_L05_Rp2}, top and middle panel).

\begin{figure*}
\includegraphics[width=0.8\textwidth]{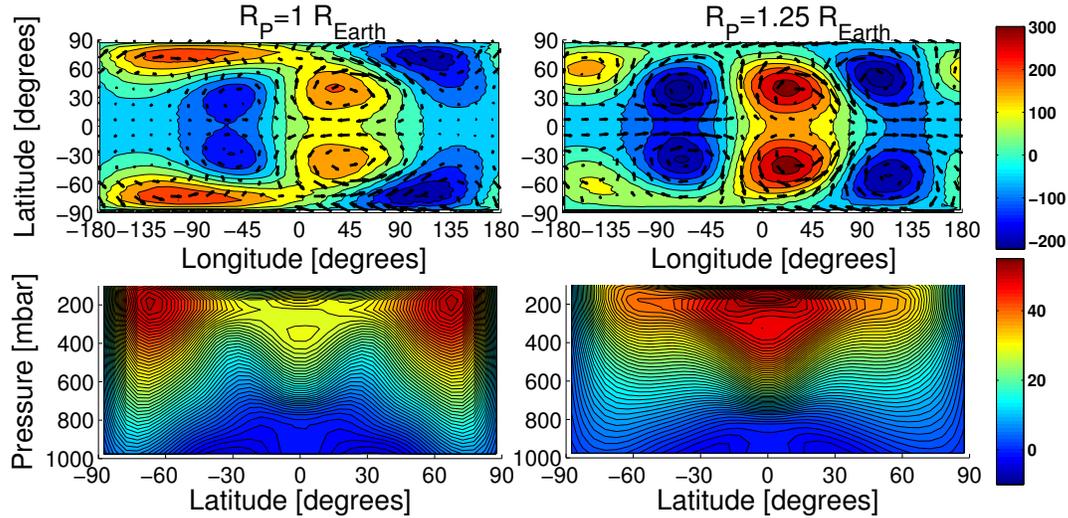}
\caption{Eddy geopotential height in [m] and eddy horizontal winds in [m/s] at p = 225 mbar (top panels) and zonal mean of zonal winds in [m/s] (bottom panels) for $R_P=1 R_{Earth}$ (left) and $1.25 R_{Earth}$ (right) and rotation period $P_{rot}=3$~day. The longest wind vectors are from left to right: 39.6, and 37.4~m/s, respectively. Contour intervals are 50~m for the eddy geopotential and 1 m/s for zonal wind speeds, respectively.}
\label{fig: Prot3d_dyn}
\end{figure*}

\begin{figure*}
\includegraphics[width=0.8\textwidth]{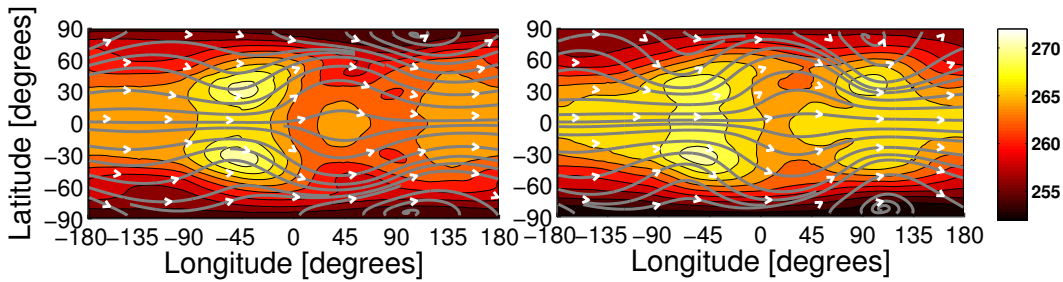}
\caption{Temperatures and streamlines of the horizontal flow, averaged over 1000 d, for pressure level p = 225 mbar for $R_P=1 R_{Earth}$ (left) and $1.25 R_{Earth}$ (right) and rotation period $P_{rot}=3$~days. Contour intervals for temperatures are 2~K.}
\label{fig: p225mb_Prot3d}
\end{figure*}

For even faster rotation, $P_{rot}=3$~day (Figure~\ref{fig: L05_Rp2} top panel)), the equatorial standing Rossby waves become again dominant, although remnants of the extratropical Rossby wave remain visible in the eddy geopotential wind field: We still find positive high latitude geopotential anomalies between longitudes $-180^{\circ}$ to -135${}^{\circ}$ and negative geopotential anomalies east from the substellar point between longitudes $+60^{\circ}$ to +135${}^{\circ}$ that appear to supersede the negative geopotential anomaly associated with the tropical Rossby wave (between longitudes $-45$ to +135${}^{\circ}$). We also find in the horizontal flow that the longitudinal location of the vortices is flipped with respect to the substellar point (Compare Figure~\ref{fig: p225mb_L05_Rp2}, top and middle panel with Figure~\ref{fig: Rossby_sketch}): They are still located at $-135^{\circ}$ for $P_{rot}=6$~days and are shifted by the appearance of the extra tropical Rossby eave to longitudes between $+90^{\circ}$ to $+135^{\circ}$ for $P_{rot}=4$~and 3 days, respectively.

Strikingly, the form and meridional tilt of the Rossby wave gyres between $-45^{\circ}$ and $135^{\circ}$ longitude are \textbf{exactly} as predicted by \cite{Showman2011} (Compare their schematic Figure~1 with Figure~\ref{fig: L05_Rp2} top-left panel). As predicted, these tilted standing Rossby waves lead to a very efficient spin-up of equatorial superrotation for $P_{rot}\leq 3$~days (E.g., Figure~\ref{fig: zonal_wind}). It will also be shown in the subsequent section that remnants of a weak standing extra tropical Rossby wave can be found, even for very fast rotating planets with strong equatorial superrotation.

Interestingly, the simulations of \cite{Edson2011} don't go into a dynamics regime that is fully dominated by standing tropical Rossby waves for $P_{rot}\leq 4$~days (See their Figures~4a,c,e and Figures~5a,c,e). Instead, their $P_{rot}=4$ and 3~days simulation are dominated apparently by standing extra tropical Rossby waves and correspondingly by $60^{\circ}$ high latitude jets and a pair of very weak jets near the equator. Their $P_{rot}=1$~day simulation shows a mixture of extra tropical and tropical Rossby wave gyres. Again, these Rossby waves drive two zonal jets per hemisphere: one at the equator and one at high latitudes. Because \cite{Edson2011} never find a purely superrotating state, the authors report maximum wind speeds of at most 111~m/s. \cite{Merlis2010} have no equatorial superrotation in their $P_{rot}=1$~day simulation and find consequently even slower wind speeds of about 50~m/s. In contrast to that, our simulations reach 300~m/s for $P_{rot}=1$~day and all planet sizes (Figure~\ref{fig: zonal_Prot1d}).

As noted in the previous section, \cite{Edson2011} identified an abrupt transition in zonal wind speeds and at least two dynamical state solutions between $P_{rot}=$3-4 days in their dry simulation. The authors tentatively linked the rotation region of multiple solutions to a $L_R/R_P=1$ transition. Now we can explain why we didn't see such an abrupt transition in zonal wind speeds and also go into a different climate state for faster rotations: The extra tropical Rossby wave is stronger than tropical Rossby waves in the model of \cite{Edson2011}. In our model, the equatorial Rossby wave is dominant. The dominance of extra tropical Rossby waves in the study of \cite{Edson2011} also explains the abrupt transition in zonal wind speeds and structure - from equatorial to high latitude - as soon as the meridional extent of the extratropical Rossby wave 'fits' on the planet and allows to form standing waves for $P_{rot}\approx 4$~days. In our simulation, equatorial Rossby waves remain dominant after transition. The extra tropical Rossby waves are suppressed until both $\lambda/R_P$ and $L_R/R_P$ become both smaller than 0.5 and both waves can 'fit' on the planet. However, even in this case the tropical Rossby wave is dominant most of the time.

\begin{figure*}
\includegraphics[width=0.8\textwidth]{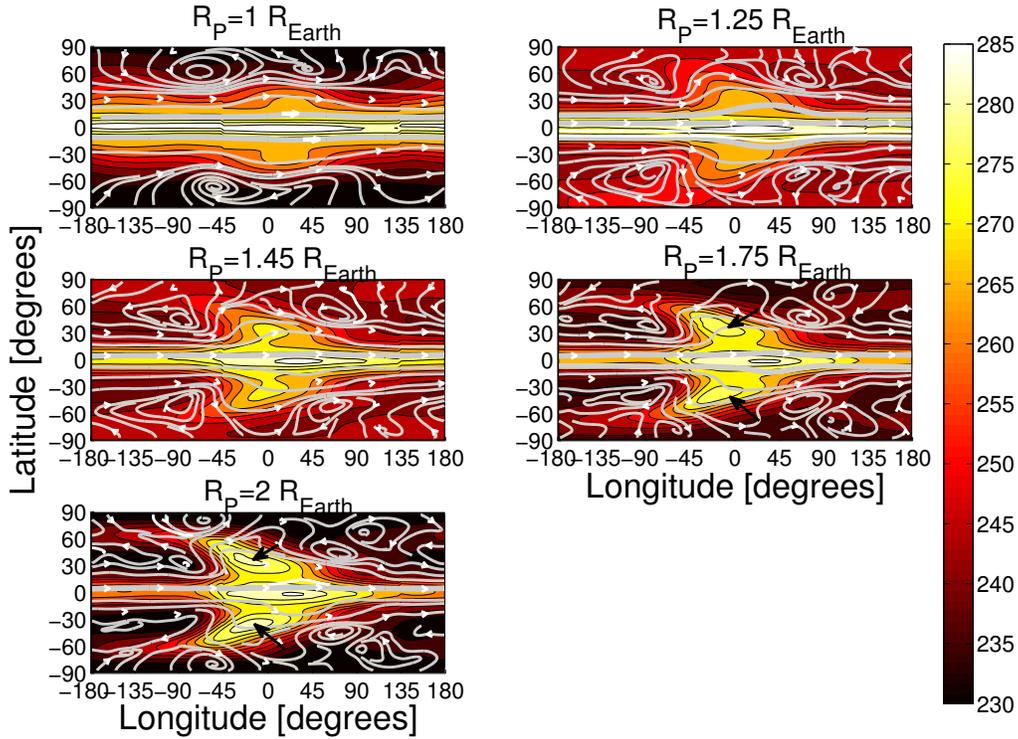}
\caption{Temperatures and streamlines of the horizontal flow, averaged over 1000 d, for pressure level p = 225 mbar for $R_P=1 - 2 R_{Earth}$ and rotation period $P_{rot}=1~day$. Contour intervals for temperatures are 5~K. The black arrows denote the hot temperature chevron shape.}
\label{fig: p225mb_Prot1d}
\end{figure*}

\begin{figure*}
\includegraphics[width=0.8\textwidth]{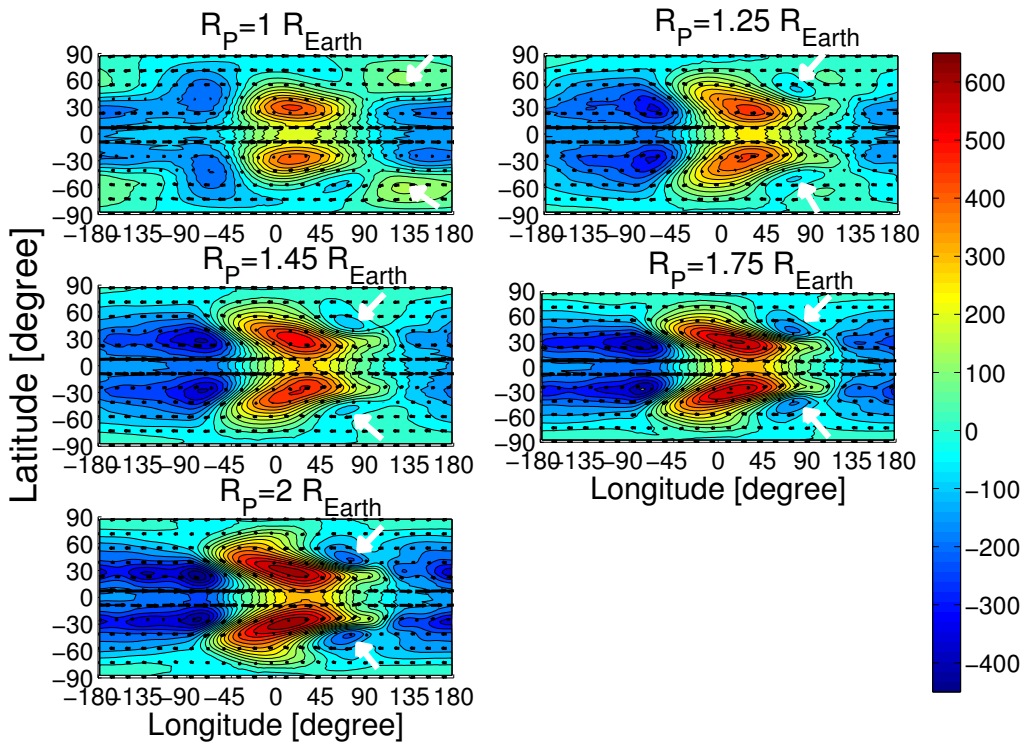}
\caption{Eddy geopotential height in [m] and eddy horizontal wind in [m/s] at p = 225 mbar for $R_P=1 - 2 R_{Earth}$ and rotation period $P_{rot}=1$~day. The longest wind vectors are from left to right and from top to bottom: 110.1, 102.0, 90.0, 95.2 and 108.4~m/s, respectively. Contour intervals are 50~m. White arrows mark extratropical eddy geopotential height anomalies.}
\label{fig: Eddy_Prot1d}
\end{figure*}

Thus, we conclude that there are two possible climate states in the $0.5\leq L_R/R_P \leq 1$ regime: One that is dominated by standing tropical planetary waves with equatorial fast zonal jets - as in our simulation - and one that is dominated by standing extra tropical planetary waves with slower high latitude jets - as shown in \cite{Edson2011}. We assume that the multiple equilibria mentioned by \cite{Edson2011} refer to these two states.

The picture becomes even more complicated for $L_R/R_P \leqslant 0.5$, where we conclude that there are now at least three dynamical states possible: a state dominated by tropical Rossby waves,  a state dominated by extra tropical Rossby waves, and a mixture of both type of waves. As soon as there is an extra tropical wave component, the zonal wind is slower compared to the purely tropical Rossby wave state.

The existence of different climate states makes the comparison of zonal wind speeds in the $L_R/R_P \leqslant 0.5$ rotation regime difficult and also explains the previously identified spread in zonal wind maxima (Figure~\ref{fig: zonal_wind}): Larger planets don't always have higher wind speeds than smaller planets and there appear to be 'jumps' in the evolution of zonal wind velocity with faster rotation. Only climates with strong equatorial Rossby waves experience very fast winds as the angular momentum is transported from high latitudes all the way towards the equator. Extra tropical waves divert at least part of the angular momentum towards $60^{\circ}$ latitude. Thus, climate states with high latitude jets are always slower than fully formed equatorial jets. Because small planets tend to have smaller wind speeds at the same rotation period than bigger planets, Earth-size planets that exhibit a mixture of tropical and extra tropical Rossby waves should show particular low wind speeds. Indeed, the slowest zonal winds in the $L_R/R_P\approx 0.5$-regime with $u_{max}\approx 50$~m/s can be found for the smallest planets in our study, $R_P=1$ and $R_P=1.25 R_{Earth}$, where both simulations show a mixture between equatorial and extra tropical wave dynamics (Compared Figures~\ref{fig: Prot3d_dyn} and \ref{fig: p225mb_Prot3d} with Figure~\ref{fig: Rossby_sketch}).

\subsubsection{Very fast rotation: $P_{rot}=1$~day.}

As already shown in Figure~\ref{fig: zonal_Prot1d}, our simulations show for all planet sizes strong equatorial superrotation. Simulations of tidally locked large exoplanets, that is, hot Jupiters (e.g. \cite{Kataria2015}) and even larger Super-Earth planets than investigated in this study (e.g. for $R_P=2.75 R_{Earth}$, \cite{Zalucha2013}) show  anti-rotating, that is, easterly jets at higher latitudes. Indeed, the two largest planets in this study, $R_P=1.75 R_{Earth}$ and $R_P=2 R_{Earth}$ show signs of anti-rotating jet formation at the surface that become more pronounced with planet size.

According to \citet{Showman2011}, the latitudinal width of an equatorial superrotating jet should correspond to the equatorial Rossby radius of deformation. This is $\pm 20$~degree for an Earth-sized planet and $\pm 13$~degree for a planet twice the size that both rotate with $P_{rot}=1$~day (Figure~\ref{fig: Rossby_radius}, left panel and Table~\ref{tab: Rossby jet width}). On the one hand, there is indeed a stronger confinement of the equatorial jet apparent for larger planets (Figure~\ref{fig: zonal_Prot1d} and Figure~\ref{fig: p225mb_Prot1d}). On the other hand, the jet widths are consistently broader than expected if the equatorial jet is formed by a standing tropical Rossby wave alone: The jet width at the top of the atmosphere exhibit a latitudinal extent of $\pm 45$ and $\pm 20$ degree for $R_P=1$ and $R_P=2 R_{Earth}$, respectively (from visual inspection of Figure~\ref{fig: p225mb_Prot1d}). Also \cite{Zalucha2013} noted that the equatorial jet width is larger in their simulations than predicted by \cite{Showman2011}. Inspection of the eddy geopotential height shows that there are, indeed, still components of extra tropical wave activity - although equatorial Rossby wave gyres dominate the flow dynamics (Figure~\ref{fig: Eddy_Prot1d}, white arrows). Apparently, the positive extra tropical and equatorial Rossby wave gyres merge around the substellar point for $R_P=1.25-2 R_{Earth}$. The merging of equatorial and extratropical Rossby waves in the strongly superrotating case explains why the equatorial jet width is broader than can be accounted for by just assuming equatorial Rossby radius of deformation. If we assume instead of $\pm \lambda_R$, $\pm \left(\lambda_R+L_R \right)$ as the latitudinal extent of the equatorial jet, we reach a better agreement (Table~\ref{tab: Rossby jet width}). Thus, it appears as if the westerly equatorial jet consists in fact of \textbf{three} westerly jets - one at the equator and one at each poleward flank, respectively. This interpretation is further supported by closer inspection of the zonal wind structure, in particular for the largest planets investigated in this study (Figure~\ref{fig: zonal_Prot1d}): There is a 'kernel' of high wind speeds between $u=$100-300~m/s at the equator with very confined latitudinal extent. This kernel is flanked at each side by a broader region with slower winds ($u=0-100$~m/s).

\begin{table}
\caption{Latitudinal extent of Rossby radius of deformation for $P_{rot}=1$~day in comparison with real width of equatorial jet.}
\begin{tabular}{c|c|c|c}
\hline
$R_P$ & $\lambda_R$ & $L_R$ & eq. jet width\\
{[}$R_{Earth}$] & [degree] & [degree]  & [degree]\\
\hline
1 & $\pm$ 20.3& $\pm$ 16.7 & $\pm$ 35-45 \\
1.25 & $\pm$ 17.3 & $\pm$ 11.2 & $\pm$ 30-40\\
1.45 & $\pm$ 15.3 & $\pm$ 9.3 & $\approx \pm$~30\\
1.75 & $\pm$ 14.3& $\pm$ 8.0 & $\approx \pm$~25\\
2 & $\pm$ 13.7 & $\pm$7.2 & $\approx \pm$~20 \\
\hline
\end{tabular}
\label{tab: Rossby jet width}
\end{table}

\section{Circulation with respect to planet size and rotation period}
\label{sec: circulation}

\begin{figure*}
\includegraphics[width=0.48\textwidth]{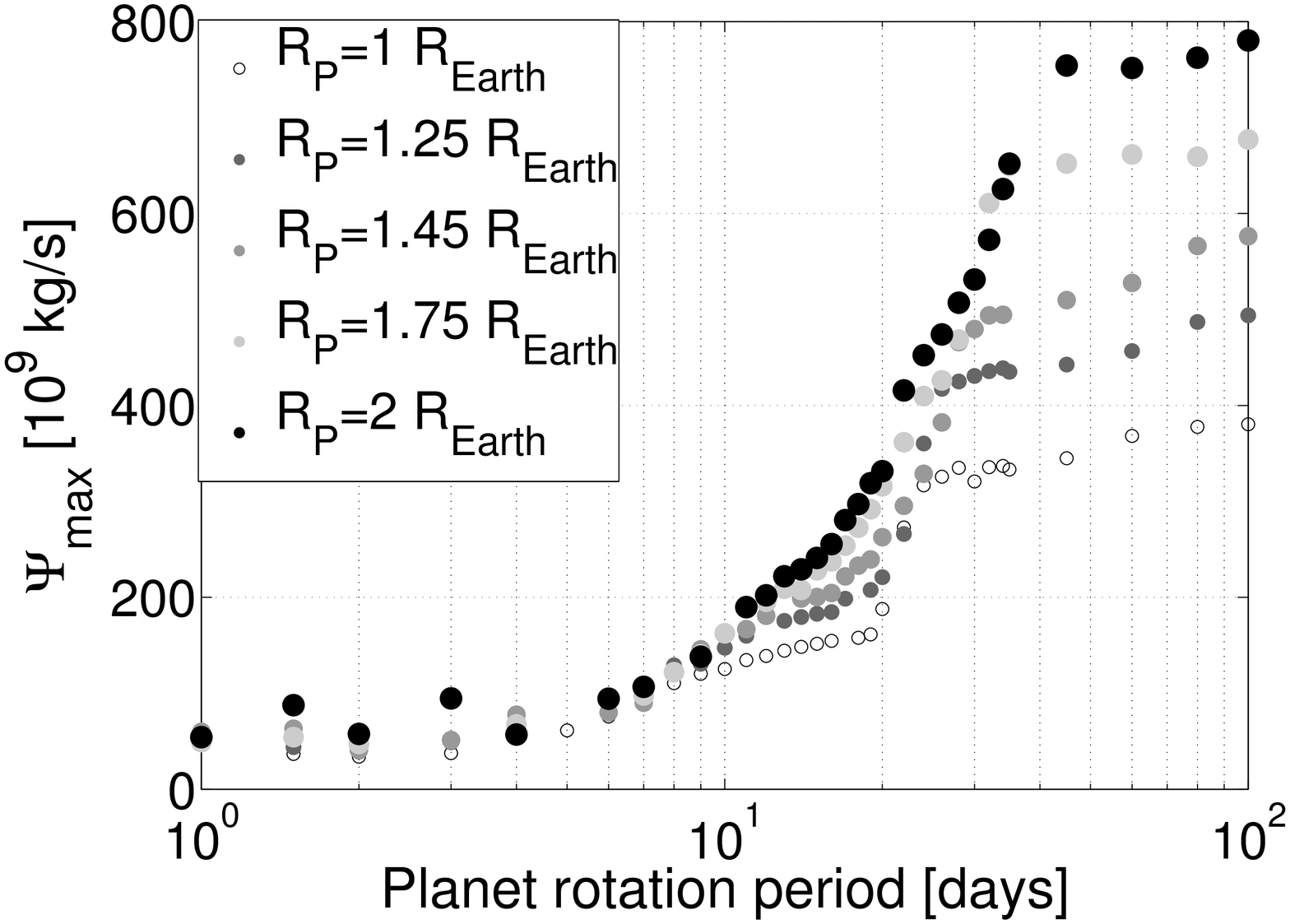}
\includegraphics[width=0.48\textwidth]{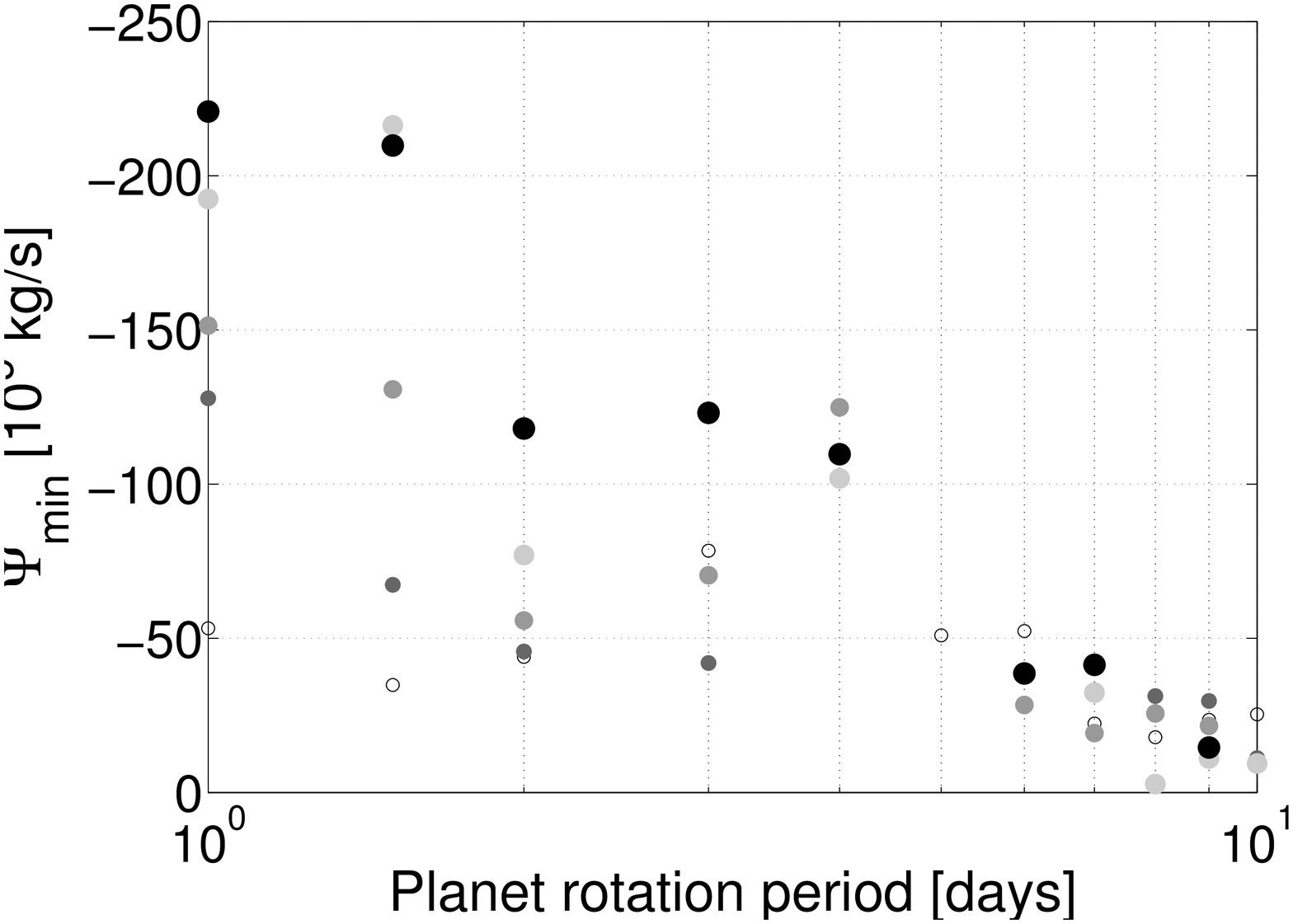}
\caption{\textbf{Left panel:} Maximum strength of the clockwise equatorial circulation cell on the North hemisphere \textbf{Right panel:} Maximum strength of a counter-clockwise secondary circulation cell on the North hemisphere.
The strength of the equatorial circulation cell is shown for the full rotation regime ($P_{rot}=1-100$~days), the strength of the secondary cell is shown for short rotation periods ($P_{rot}=1-10$~days) of a tidally locked terrestrial planet for different planet radii ($R_P=1-2 R_{Earth}$), assuming moderate Earth-like irradiation. }
\label{fig: Psicomp}
\end{figure*}

In the context of habitability, circulation is a dynamical features of uttermost importance: Circulation determines regions of cloud formation and of high and low precipitation. First insights into the change of circulation regime with rotation period was already found in \cite{Carone2014}. The study showed that a slow rotating tidally locked Super-Earth planet with $P_{rot}=36.5$~days exhibits two strong direct circulation cells - one for each hemisphere - with an uprising branch at the substellar point that transports hot air towards the night side across the poles and terminator. This finding is in agreement with the majority of circulation studies on terrestrial planets (e.g. \cite{Merlis2010}, \cite{Navarra2002}). For $P_{rot}=10$~days, the circulation was found to be more complex: The atmosphere still exhibits two large global circulation cells but also contains two smaller embedded counter-circulating cells in the mid-atmosphere at $p=$500~mbar. \cite{Joshi1997} showed a very similar circulation structure for $P_{rot}=16$~days. \cite{Edson2011}, on the other hand, did not report this circulation state. Interestingly, even between complex Earth models, there is a disagreement of the circulation state for fast rotating tidally locked planets. \cite{Edson2011} find three circulation cells per hemisphere for $P_{rot}=1$~day, where the equatorial direct cell is much stronger than the other two circulation cells. \cite{Merlis2010} report for the same rotation period also three circulation cells per hemisphere. But the polar circulation cells are the strongest in their simulations.

\begin{figure*}
\includegraphics[width=0.9\textwidth]{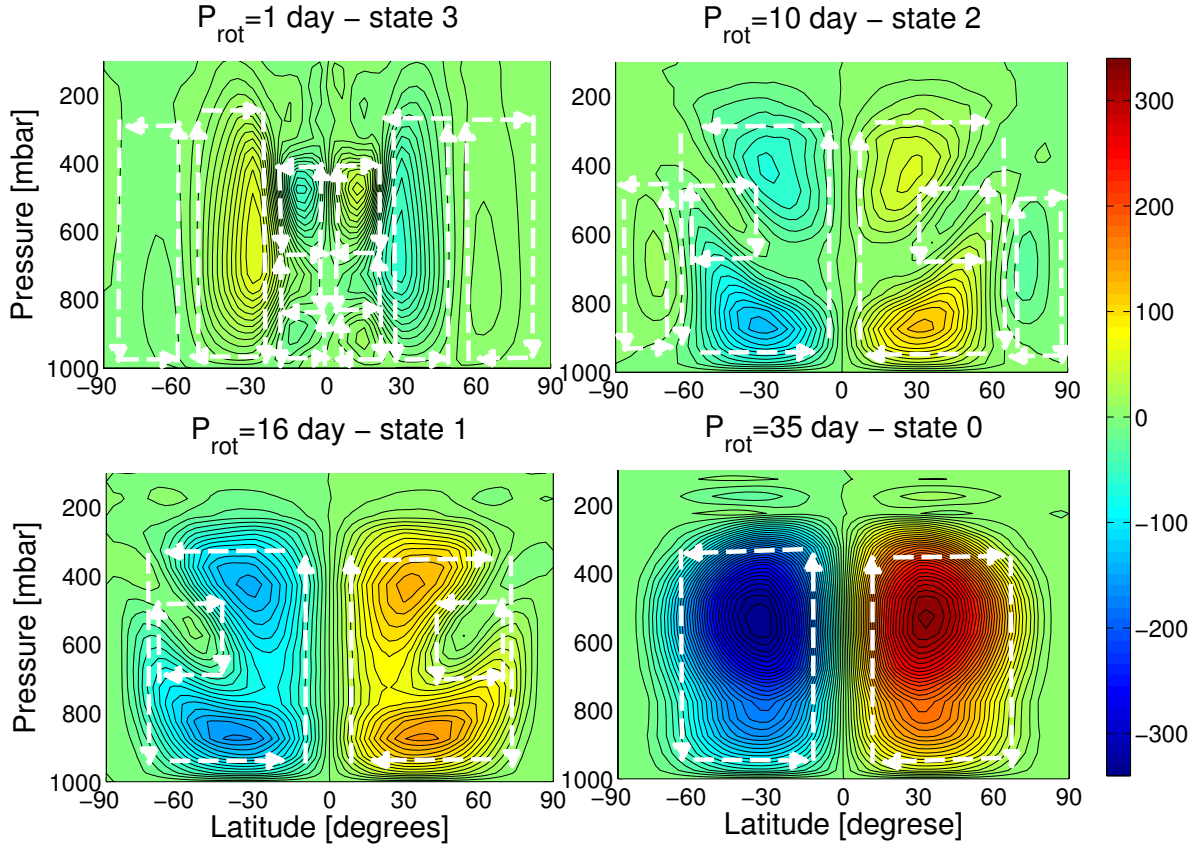}
\caption{Meridional mass flux stream function $\Psi$ in units of $10^9$ kg/s for $R_P=1 R_{Earth}$ showing the four possible circulation states and the general reduction of circulation strength with faster rotation. Note that positive values indicate clockwise and negative values counterclockwise circulation. Contour intervals are $5\times 10^9$ kg/s for $P_{rot}=1$~day and $10\times 10^9$ kg/s elsewhere. White dashed arrows mark the direction of circulation.}
\label{fig: circulation_state}
\end{figure*}

In the following, we will precisely identify transition between different circulation states with respect to rotation period and planet size. Furthermore, we will discuss possible implication for habitability. Henceforth, we call the circulation state with one unperturbed global circulation cell per hemisphere that is found for slow rotators \textit{state~0}. This study allows to exactly locate the rotation periods where the embedded reverse circulation cells appear (\textit{state~1}) and also to check if and when two or more large scale circulation cells emerge. We call a circulation state with two fully formed circulation cells per hemisphere \textit{state 2} and circulation states with three and more circulation cells per hemisphere \textit{state 3}. The rotation periods of regime transitions are shown in Table~\ref{tab: circulation}. The circulation states and directions of circulation are shown for an Earth-size planet in Figure~\ref{fig: circulation_state}.
\begin{table}
\caption{Circulation state transitions for different planet sizes
in our nominal model.}
\begin{tabular}{c|c|c|c|c}
\hline
 & state 0 &state 1 & state 2 & state 3\\
\hline
$R_P$ & $P_{rot}$&$P_{rot}$ & $P_{rot}$ & $P_{rot}$\\
{[}$R_{Earth}$]&[days] & [days] & [days]  & [days]\\
\hline
1 &100 - 22& 20 - 13& 12 - 1.5& 1 \\
1.25 &100 - 24& 22 -12 & 11 - 2 & 1.5 - 1\\
1.45 & 100 - 26& 24 - 14& 13 - 3 & 2 - 1\\
1.75 & 100 - 30& 28 - 10& 9 - 4 & 3 - 1\\
2 & 100 - 34&32 - 10& 9 - 5&4 - 2\\
\hline
\end{tabular}
\label{tab: circulation}
\end{table}

The appearance of an embedded circulation cell \textit{(state~1}) can already be identified by monitoring the maximum and minimum of the meridional mass transport stream function $\Psi$. This is defined as
\begin{equation}
\Psi=\frac{2\pi R_P}{g}\cos\nu \int_0^p \bar{v} dp',
\end{equation}
where $\nu$ is latitude and $\bar{v}$ is the zonal and temporal mean of the meridional velocity component $v$ at a given latitude\footnote{Note that the horizontal wind velocity is $\vec{v}=(u,v)$, where $u$ is the zonal and $v$ the meridional component.}. $\Psi$ is positive for clockwise circulation and negative for counter-clockwise circulation.

The transition from \emph{state~0} to \emph{state~1} corresponds to a steep decrease in the maximum meridional mass stream function $\Psi_{max}$ (Figure~\ref{fig: Psicomp}, left panel). It sets in after the tropical Rossby wave transition, $\lambda_R/R_P=1$ (Table~\ref{tab: Rossby wave} and Figure~\ref{fig: Rossby_radius}). Apparently, the embedded cells appear when the tropical Rossby wave can form a standing wave. The connection between tropical Rossby waves and embedded cell becomes even stronger when taking into account the findings in \cite{Carone2014}: The embedded reverse circulation cells in the meridional mass transport function appear to be latitudinal cross sections of Walker-like circulation cells. They were shown to transport air masses primarily along the equator. Indications of this longitudinal circulation could also be found in \cite{Joshi1997} that also report direct meridional circulation cells with embedded reverse cells. A climate state with tropical Rossby waves and Walker-circulation can be found on Earth in the tropical regions and is driven by longitudinal differences in surface heating \citep{Gill1980}. Therefore, we see in the Earth's tropics on a small scale what tidally locked terrestrial planets experience on a global scale \citep{Showman2011}. The embedded reverse circulation cells, once formed, are present even for faster rotations (Fig.~\ref{fig: circulation_state}).

The absence of a \emph{state~1} circulation cell in the model of \cite{Edson2011} and the presence of a second extended circulation cell per hemisphere already for $P_{rot}=10$~days can be explained by the dominance of the extra tropical Rossby waves in their model. On Earth, the second pair of circulation cells, the Ferrell cells, are located polewards of the direct tropical circulation cells and are a result of eddy circulation and are flanked by extra tropical Rossby waves (the subtropical and tropical jet, see, e.g. \cite{Holton}). Thus, we speculate that the second pair of circulation cells per hemisphere for the planets with $P_{rot} \geq 10$~days shown in \cite{Edson2011} are likewise driven by eddy circulation and are associated with strong extra tropical Rossby waves that are stronger in their model than in our nominal model. The relative weakness of extra tropical Rossy waves in our model may also explain why the secondary circulation cells (\textit{state 2}), polewards of the direct equatorial circulation cell, are much weaker in the $P_{rot}=10$~days simulation of an Earth-size planet (Fig.~\ref{fig: circulation_state}) than for the same set of planet parameters in \cite{Edson2011} (their Fig.~4i).

While for the transition from \textit{state 0} to \textit{state 1}, a link can be found between circulation and standing (tropical) Rossby waves, no such link can be established at first glance for the other circulation state transitions with rotation period when inspecting Table~\ref{tab: circulation}. This is surprising because the $L_R/R_P \approx 1$ and $\lambda_R/R_P \approx 0.5$ transition regions were previously identified with suppression of the uprising direct circulation branch (Section~\ref{sec: L/R-1}). Indeed, the equatorial direct circulation cell is very weak and also vertically suppressed in \textit{state 3}. This state, however, is only established for relatively fast rotation ($P_{rot} \leq 4$~days).

\begin{figure*}
\includegraphics[width=0.9\textwidth]{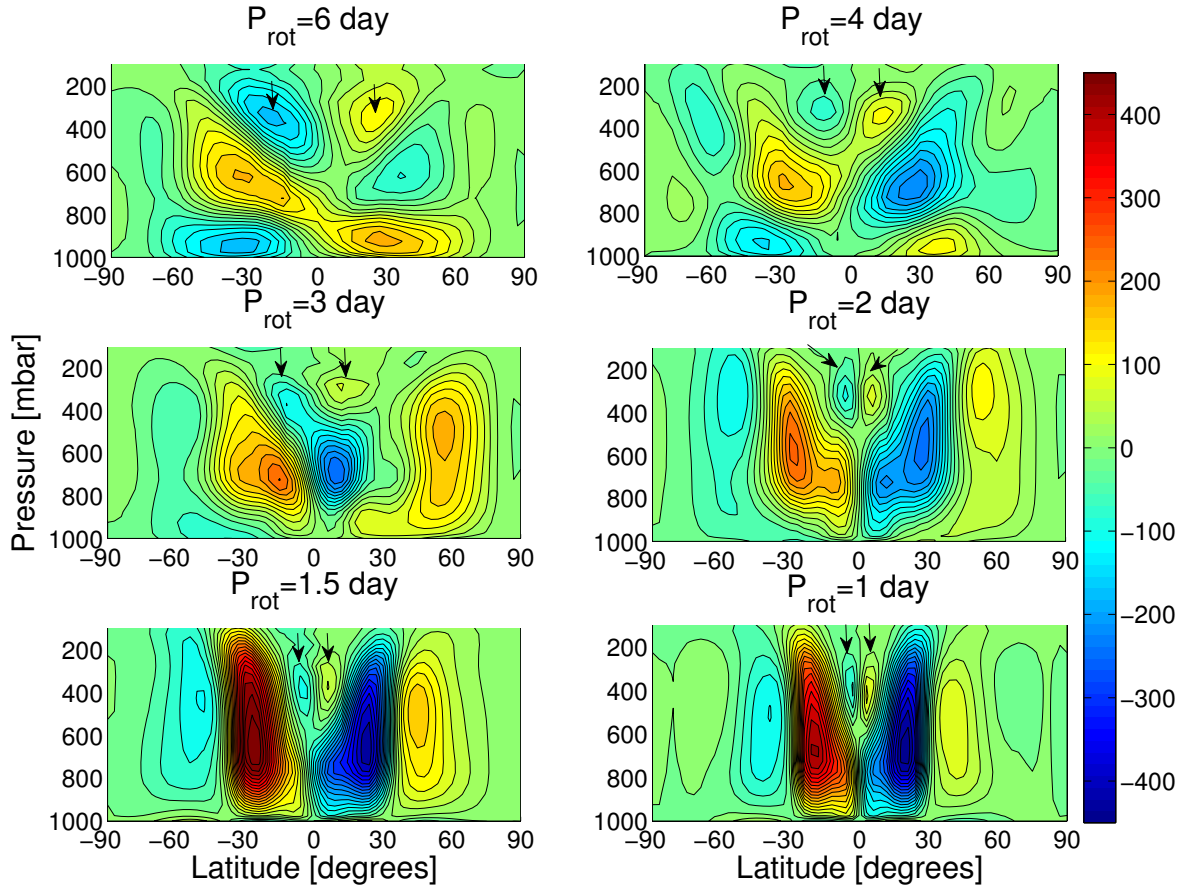}
\caption{Meridional mass flux stream function $\Psi$ in units of $10^9$ kg/s for $R_P=2 R_{Earth}$ showing the four possible circulation states and the general reduction of circulation strength with faster rotation. Note that positive values indicate clockwise and negative values counterclockwise circulation. Contour intervals are $25\times 10^9$ kg/s. Arrows indicate the upper segment of the direct circulation cells.}
\label{fig: circulation_2Rp}
\end{figure*}

A closer inspection of Figure~\ref{fig: circulation_2Rp}, top-left panel, shows for the $P_{rot}=6$~days simulation is illuminating: Although the direct equatorial circulation cell is not fully suppressed, it appears to be fragmented into a lower segment close to the surface between $p=1000-800$~mbar and an upper segment at $p=500-100$~mbar by the embedded circulation cell. The disappearance of upwelling signatures at the top of the atmosphere over the substellar point in Section~\ref{sec: L/R-1}, thus, would not indicate the complete disappearance of direct circulation but instead its fragmentation. As a consequence, only a part of rising air masses over the substellar point is carried high aloft, whereas the other part is transported close to the surface. Condensation of volatile species at high altitudes after evaporation at the surface, and thus cloud formation, may be suppressed in such a circulation state. We find the fragmentation to be the strongest for the $R_P=2 R_{Earth}$ planet (Figure~\ref{fig: circulation_2Rp}). It makes the identification of distinct circulation cells very difficult. It is doubtful if a closed transport cycle of condensables, like the water cycle, can be established in such fragmented circulation states.

At the very least, the onset of circulation \textit{state 2} with a secondary full circulation cell per hemisphere should depend on rotation period and planet size, if one naively assumes that secondary circulation cells form once the direct circulation cells have shrunk sufficiently in latitudinal extent. According to \cite{Held1980}, the latitudinal extent of the direct cell is
\begin{equation}
\nu_H=\left(\frac{5 g H \Delta \theta_{p-eq} }{3 \Omega^2 R_P^2}\right)^{1/2} \label{eq: Hadley},
\end{equation}
where $ \Delta \theta_{p-eq}$ is the fractional change in potential temperature from pole to equator. This relation predicts that the extent of the direct Hadley circulation cell shrinks with planet size and angular velocity. Thus, one would expect a transition to \textit{state 2} with more circulation cells to appear at slower rotations for a large planet when compared to a small planet.

Instead, we find that the transition to \textit{state 2} is shifted to smaller rotation periods with increasing planet size (Table~\ref{tab: circulation}). Equation~\ref{eq: Hadley} also predicts that the Hadley cell extent $\nu_H$ is smaller than $90^{\circ}$ already for $P_{rot}=100$~days and for any planet size. According to these results, we should not see a \textit{state 0} circulation state in our parameter study. However, Equation~\ref{eq: Hadley} is only truly applicable to direct unperturbed circulation cells, only for small Coriolis forces and was developed to describe the circulation regime of the Earth, where the temperature gradient is between equator and pole and not between day side and night side. It is, thus, not really surprising that this simple relation fails to capture the evolution of circulation cells in our study. 

Direct circulation cells not only lose in strength with fast planet rotation and become fragmented, the cells can become almost fully suppressed in circulation \textit{state 3} (Fig.~\ref{fig: circulation_2Rp}, lowest panels, and \ref{fig: circulation_state}, top-left panel): They appear 'pushed' down towards the surface by the equatorial superrotation.

The suppression of the direct circulation starts at $P_{rot}\approx 4$~days for all planet sizes (Fig.~\ref{fig: Psicomp}, left panel), whereas secondary circulation cells gather strength and even start to dominate (Fig.~\ref{fig: Psicomp}, right panel) for large terrestrial planets ($R_P=1.25-1.75 R_{Earth}$). See also Figure~\ref{fig: circulation_2Rp}, bottom panels, where we have added arrows for the better identification of the direct circulation cell remnants\footnote{As a reminder, direct circulation is counter-clockwise in the South hemisphere (negative latitudes) and clockwise in the North hemisphere (positive latitudes), for the secondary circulation the circulation direction is reversed (see Figure~\ref{fig: circulation_state}).}. Interestingly, the transition to circulation \textit{state 3} not only suppresses direct circulation, it also stabilizes the other circulation cells against fragmentation for $R_P=2 R_{Earth}$ (Figure~\ref{fig: circulation_2Rp}, bottom panels). A stable circulation \textit{state 3} may make the stable transport cycle of volatiles again possible. The transport, however, has to occur - in the absence of direct circulation - via the secondary and tertiary cells.

Apparently, full 3D planet atmosphere models are required to track the evolution of circulation, which is already known from the Earth climate regime \citep{Holton}. But even then we need to be wary of differences in the climate dynamics - like the relative strength of the tropical and extra tropical Rossby wave - that may lead to drastic deviations between the assumed circulation state. 

The suppression of direct circulation for fast rotating terrestrial planets that we report here is indeed very different from the results obtained by \cite{Edson2011}, where direct circulation is always dominating. Even for $P_{rot}=1$~day with a mixed state between high latitude and equatorial westerly jets, the direct circulation cells are not suppressed but flanked by two jets at $\nu\pm 10^{\circ}$. The wind speeds are, however, much slower ($u\approx 100$~m/s) than the ones found in our simulations ($u\approx 300$~m/s). \cite{Merlis2010} don't have superrotation but high latitude westerly in their $P_{rot}=1$~day simulation. Their direct equatorial and secondary circulation cells are of similar strength. All three studies are in different climate states for an Earth-sized planet with $P_{rot}=1$~days, which explains why they disagree in the details of the circulation state. At least they agree that a fast rotating tidally locked Earth-size planet has three circulation cells per hemisphere.

\section{Temperatures and habitability}
\begin{figure*}
\includegraphics[width=0.45\textwidth]{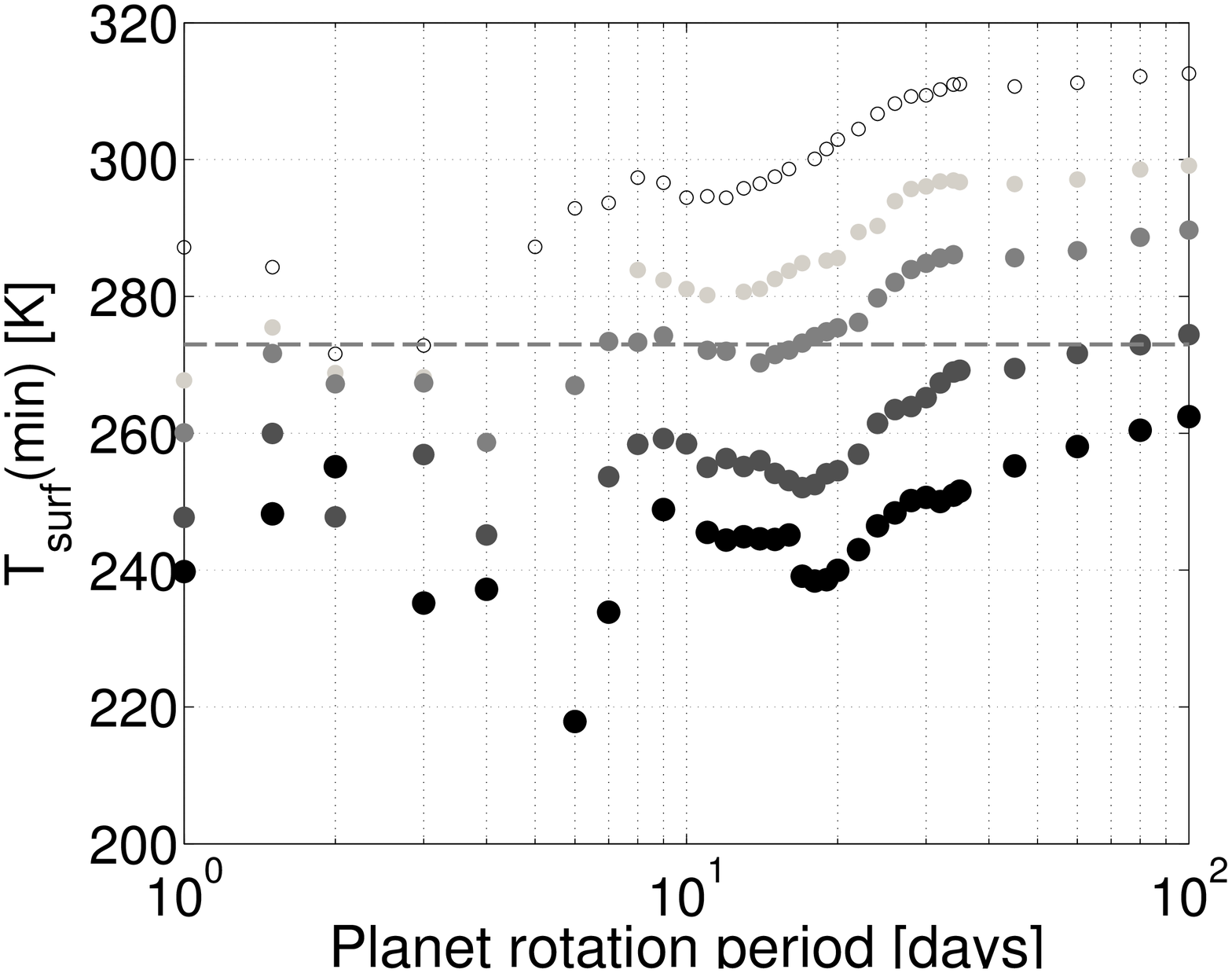}
\includegraphics[width=0.45\textwidth]{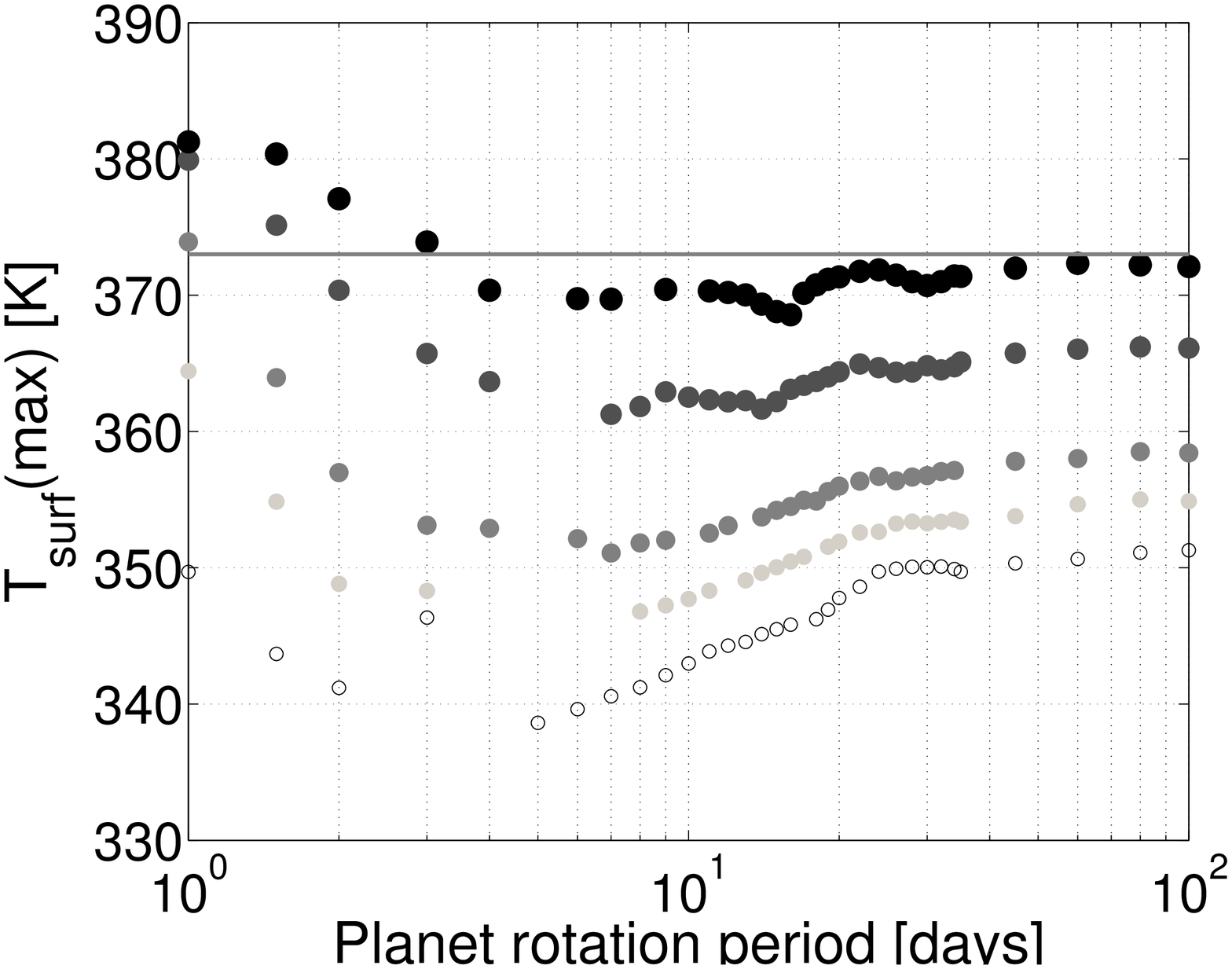}
\includegraphics[width=0.6\textwidth]{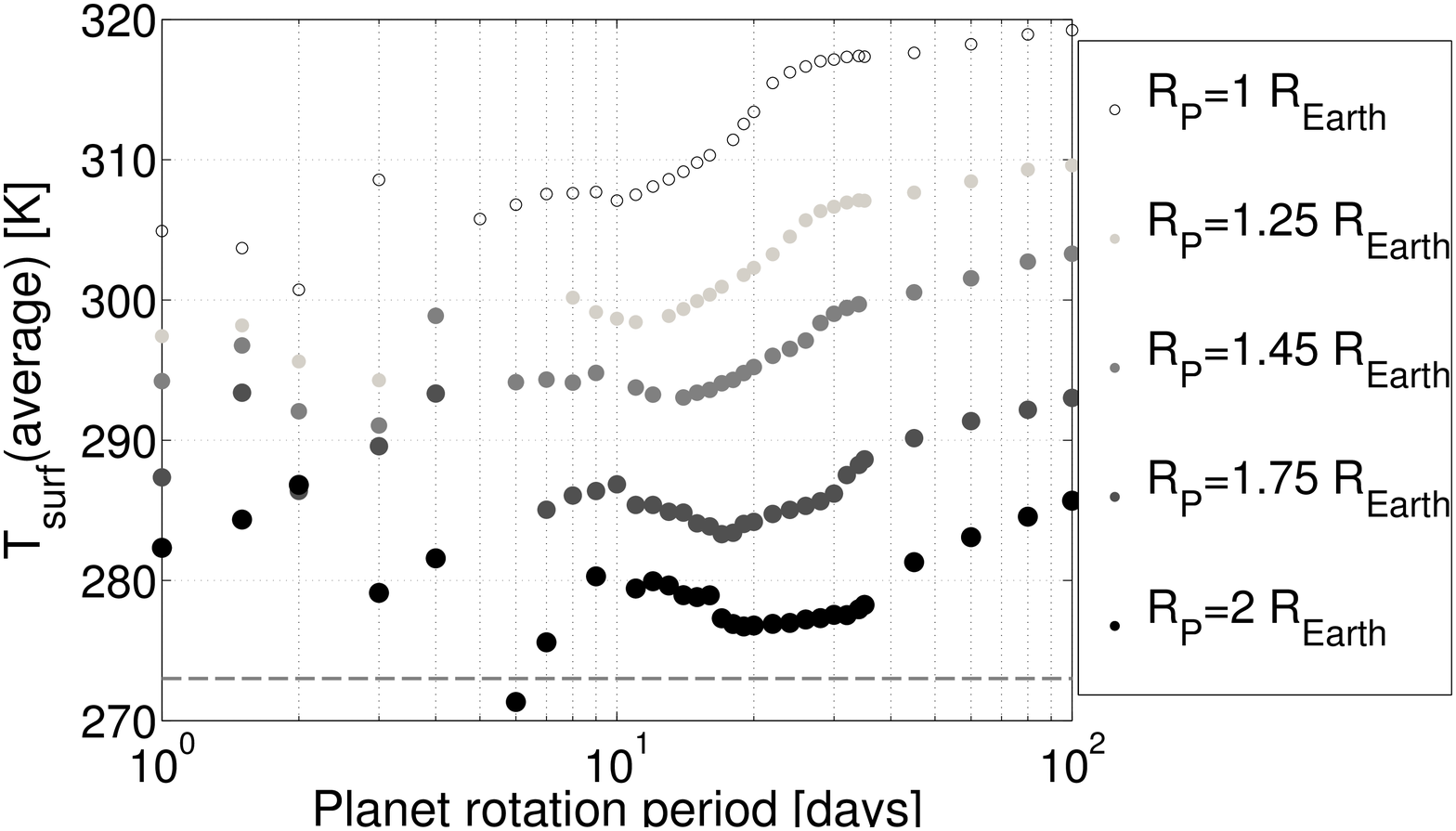}
\caption{ Minimum surface temperatures (top-left panel),  maximum surface temperatures (top-right panel) and average surface temperatures (lower panel) versus rotation period and planet size for Earth-like irradiation and an Earth-like atmosphere. The dashed gray line marks $T=273$~K, the solid gray line marks $T=373$~K.}
\label{fig: Tsurf}
\end{figure*}

Therefore, we have shown once again the merit of our detailed and coherent parameter space investigation with regards to standing Rossby waves. We can tie our improved understanding of the Rossby wave climate regimes derived in the previous section to changes in the circulation states. We can even coherently explain differences in climate states between different models.

Changes in circulation and in climate dynamics due to different standing planetary wave configurations have an influence on surface temperatures and thus, potentially, on habitability. Thus, we discuss in the following the evolution of surface temperatures with planet size and rotation period in the light of the previously gained insights.

\subsection{Surface temperature dependency on planet size}

First of all, bigger planets tend to have larger surface temperature gradients in our study: they have a hotter substellar point and cooler night sides than smaller planets (Figure~\ref{fig: Tsurf}). Furthermore, the average surface temperature is higher for small planets compared to big planets. These results are surprising as the circulation strength appears to be stronger for larger planets (Figure~\ref{fig: Psicomp}, left panel) and also the wind speeds are higher (Figure~\ref{fig: zonal_wind}). Both effects would suggest a better temperature mixing and thus a smaller temperature gradient for large planets. However, the surface temperature evolution not only depends on wind speeds and circulation, that is, on planet atmosphere dynamics, but also on the stellar irradiation. In the context of stellar irradiation, the eternal night side in our model always cools down and the eternal day side is always irradiated. The efficiency of the radiative cooling/heating with planet size can be estimated via Equation~\ref{eq:tau_rad}: The efficiency increases with increasing planet size and thus increasing surface gravity $g$. The dynamical time scale $\tau_{dyn}=U/R_P$ (Section~\ref{sec: time scale}), on the other hand, is expected to also increase with $\sim R_P$, at least for slow rotation, because we would expect $U\sim R_P^2$. Thus, the increase in radiation efficiency is naively expected to be balanced by increased dynamical efficiency. The comparison between zonal wind maxima in Section~\ref{zonal_wind_Rp} indicates that the wind speeds indeed increase as expected with planet size. The relations discussed in Section~\ref{zonal_wind_Rp}, however, are only valid for zonal wind velocities above the planetary boundary layer, which is true for the discussed maximum zonal wind speeds. For surface temperatures, however, surface friction can no longer be neglected. Furthermore, it already has been shown in Section~\ref{zonal_wind_Rp} that $U \sim R_P^2$ is only a valid assumption for slow rotation with negligible Coriolis force. Therefore, in the fast rotation regime, the increase in wind velocities with planet size is smaller than estimated by the scale analysis. For these two reasons, the dynamical efficiency of surface flow does increase with planet size but not as fast as the efficiency of radiative forcing. Consequently, larger planets have cooler night sides and hotter day sides.

\subsection{Surface temperature dependency on planet rotation period}

The dependency of surface temperatures with rotation is more complex but follows tendencies that have been identified earlier with respect to horizontal flow (Section~\ref{sec: phase transition}) and circulation (Section~\ref{sec: circulation}). Slow rotating planets ($P_{rot} = 30 -100$~days) are well mixed by the direct equatorial cell (\textit{state 0}) and, thus, show the smallest surface temperature gradient (Figure~\ref{fig: Tsurf}, upper panels). With faster rotation and thus diminishing direct circulation strength (Figure~\ref{fig: Psicomp}, left panel), the night side temperatures drop more rapidly than the substellar point temperatures, which leads generally to lower average surface temperatures (Figure~\ref{fig: Tsurf}, lower panel). Interestingly, the day side surface temperatures also decrease slightly with decreasing circulation strength. This suggests that the night side branch of the circulation that heats the night side via transport from the hot substellar point, is more strongly diminished than the day side branch that cools the substellar point via upwelling. The difference can be explained by taking into account surface friction that diminishes the surface flow and thus acts primarily on the lower branch of the circulation with flow from the night side towards the substellar point (see also \cite{Carone2014}). The longer air masses remain on the night side, the more heat can be radiated away, the cooler the air becomes. The vertical uprising branch directly over the substellar point is, on the other hand, only indirectly affected by surface friction. The net effect is an overall cooling of the surface.

The transition from a state with only direct circulation (\textit{state 0}) to a state with direct circulation and embedded reverse circulation cells (\textit{state 1}) can be identified in Figure~\ref{fig: Tsurf} by a change in the cooling rate of average surface temperatures between $P_{rot}=20 -34$~days, depending on planet size, respectively (Compare with Table~\ref{tab: circulation}). The circulation transition results in a strong decrease in minimum night side temperatures due to the reduced efficiency of heat transport from the day to the night side (Figure~\ref{fig: Psicomp}, left panel). Interestingly, the substellar point surface temperature also cools down although this effect is less pronounced for larger planets.

\textbf{\begin{figure*}
\includegraphics[width=0.9\textwidth]{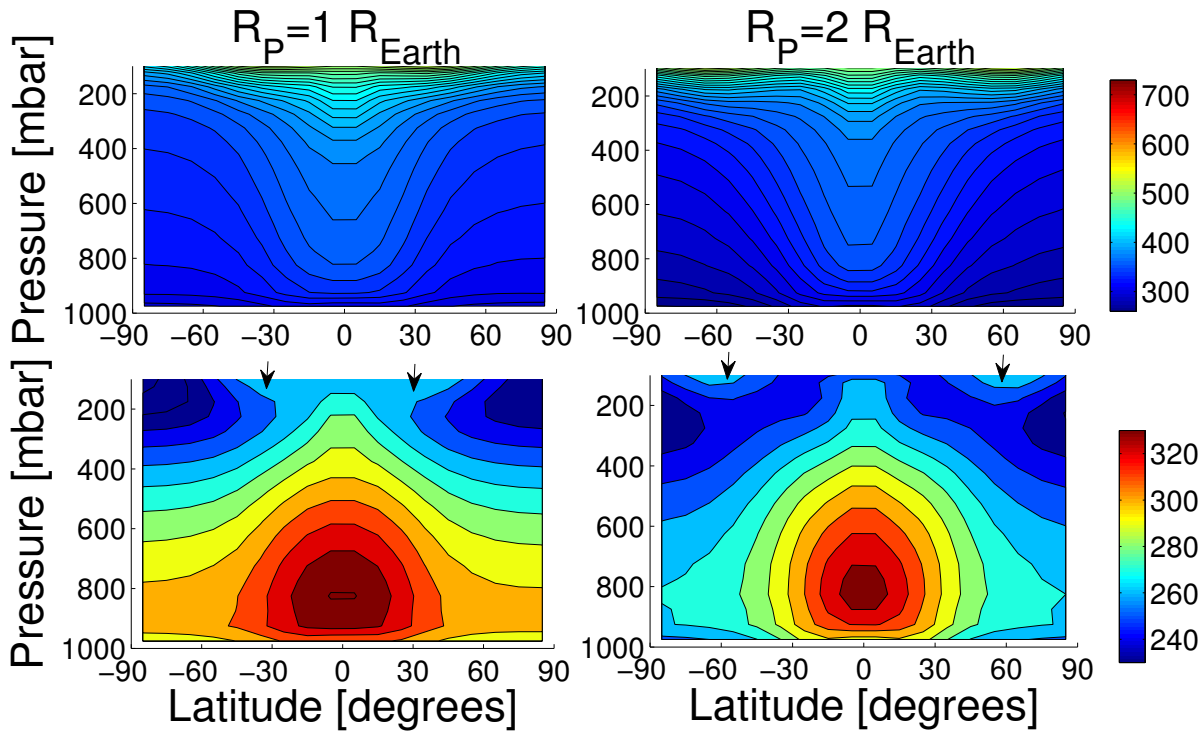}
\caption{Zonally and temporally (1000~days) averaged potential temperature (upper panels) and temperature in K for an Earth and 2 Earth size planet and rotation period $P_{rot}=1$~days. Contour intervals are 10~K. Arrows indicate location of thermal inversions.}
\label{fig: T_Prot1d}
\end{figure*}}

With even faster rotations, the investigated climate simulations approach the $L_R\approx 1$ and $\lambda_R\approx 0.5$ transition region (Section~\ref{sec: L/R-1}). We associated this region with fragmentation of the uprising branch of the direct circulation over the substellar point and thus reduced upwelling over the substellar point. At the same time, secondary circulation is still very weak. Thus, it is puzzling that we find in this region apparently a warmer night side and average surface temperatures ($P_{rot}=8-12$~days for $R_P=1 R_{Earth}$ and $P_{rot}=9-16$~days for $R_P=2 R_{Earth}$, compare Table~\ref{tab: Rossby wave} with Figure~\ref{fig: Tsurf}). The fragmentation of the direct circulation cells into a lower and an upper segment may explain the warmer night sides. We already speculated in Section~\ref{sec: circulation} that air masses are transported closer to the surface in the lower segment, which would represent a 'short cut' compared to a circulation state with a fully vertically extended circulation cell. The lower circulation segment may thus heat the night side surface efficiently. 

It is puzzling though that day side temperatures remain constant for large planets ($R_P \geq 1.75 R_{Earth}$), until the $L_R/R_P \approx 0.5$ transition with $P_{rot}\approx 4$~days is reached (Table~\ref{tab: Rossby wave}). For small planets ($R_P \leq 1.45 R_{Earth}$), the day side temperatures cool down with faster rotation, as expected, because direct circulation strength decreases with faster rotation. The days side cooling stops at the $L_R/R_P \approx 0.5$ transition region ($P_{rot}=5-6$), where the simulations reach another atmosphere dynamics regime.

The rotation regime with $L_R/R_P \leq 0.5$ brings a rise in surface temperatures for all planet sizes, where the substellar point temperatures become very hot and the night side is moderately warmed. The strong increase in day side surface temperatures can be explained by the suppression of the equatorial direct circulation cell and thus upwelling that would normally cool down the substellar point efficiently. At the same time, secondary circulation cells gain in strength and manage to transport heat from higher latitudes at the day side towards the night side - albeit not as efficient as the direct circulation in the slow rotation regime $P_{rot} \geq 30$~days. This explains the moderate warming of the night side surface.

The fragmented circulation for $P_{rot}=2-6$~days on the $2$~Earth radii planet is reflected in fluctuating night side minimum temperatures (Figure~\ref{fig: circulation_2Rp}). Apparently, the night side is even more strongly affected by this strange circulation than the day side. 

In summary, only the circulation transition from \textit{state 0} to \textit{state 1} and the fragmented circulation of the 2 Earth-radii planet correspond to changes in surface temperatures. Fragmented circulation also mainly affects the night side. Climate dynamics transitions brought about by changes in standing Rossby waves, on the other hand, play again a strong role in shaping surface temperatures via changes in the horizontal flow. We have already discussed in \cite{Carone2014} that our surface temperatures are at the night sides considerably warmer than in the complex Earth model studies of \cite{Edson2011} and \cite{Joshi2003}, while they agree roughly with \cite{Joshi1997}. The latter also used a model with simplified forcing. We will show in an upcoming follow-up study, that it is possible to change the temperature forcing prescription of \cite{Carone2014} to include more efficient night side cooling. We will discuss there how and if more efficient night side cooling changes the results derived in this study.

\subsection{Vertical temperature profile}
\label{sec: vertical temperature}
The surface temperatures are only a small part of the total temperature structure in a planet's atmosphere. In the following, we will thus discuss the complete vertical temperature structure.

\cite{Edson2011} showed that the slope of the isentropes in potential temperature between the tropics and the mid-latitudes is much gentler for a tidally locked Earth-size planet with $P_{rot}=1$~days than in the dry Earth control run. \cite{Edson2011} attributed that to a standing Rossby wave that we have identified in this study as a standing \textbf{extra tropical} Rossby wave.

The vertical potential temperature profile for a tidally locked Earth-size planet with $P_{rot}=1$~day in our study (Figure~\ref{fig: T_Prot1d}) exhibits much steeper isentropes than in Figure~4b of \cite{Edson2011}. This difference is again evidence that our simulations are in a very different climate dynamics regime than the simulations of \cite{Edson2011}: one that is dominated by standing tropical instead of extra tropical Rossby waves. Our results show, furthermore, that the isentropes are steeper for a larger planet ($R_P=2 R_{Earth}$). These results are expected as the steepness of the isentropes are also determined by the strength of the Coriolis force, which is stronger on a larger planet. Furthermore, we see evidence of temperature inversion at the top of the atmosphere for this very fast rotation, that is, a local increase in temperature with height instead of a decrease. 

\begin{figure}
\includegraphics[width=0.45\textwidth]{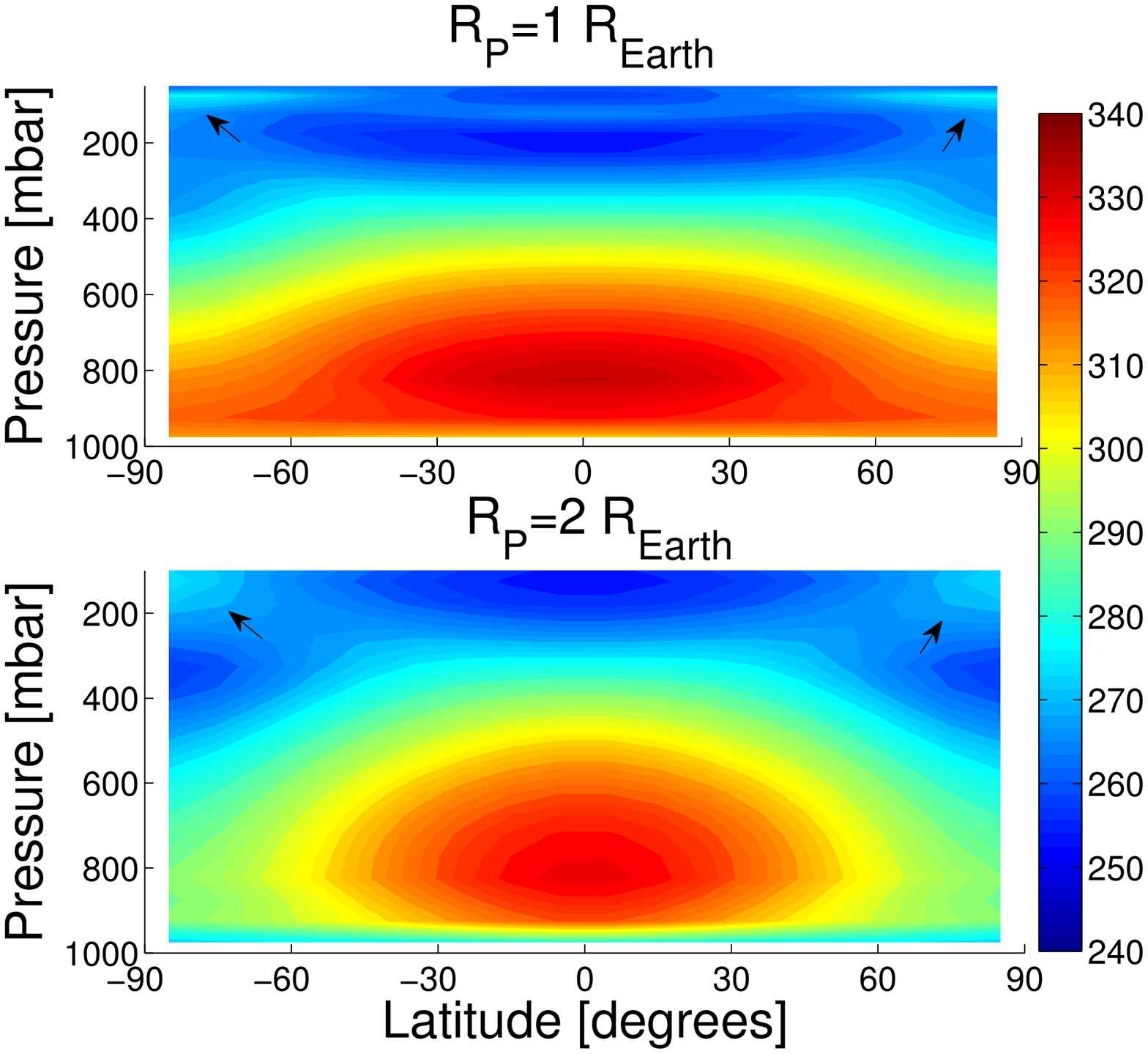}
\caption{Zonally and temporally (1000~days) averaged temperature in K for an Earth (upper panel) and 2 Earth size planet (lower panel). In both cases: $P_{rot}=16$~days. Contour intervals are 2~K. Black arrows mark the location of the upper atmosphere thermal inversion.}
\label{fig: T_Prot16d}
\end{figure}

\begin{figure*}
\includegraphics[width=0.9\textwidth]{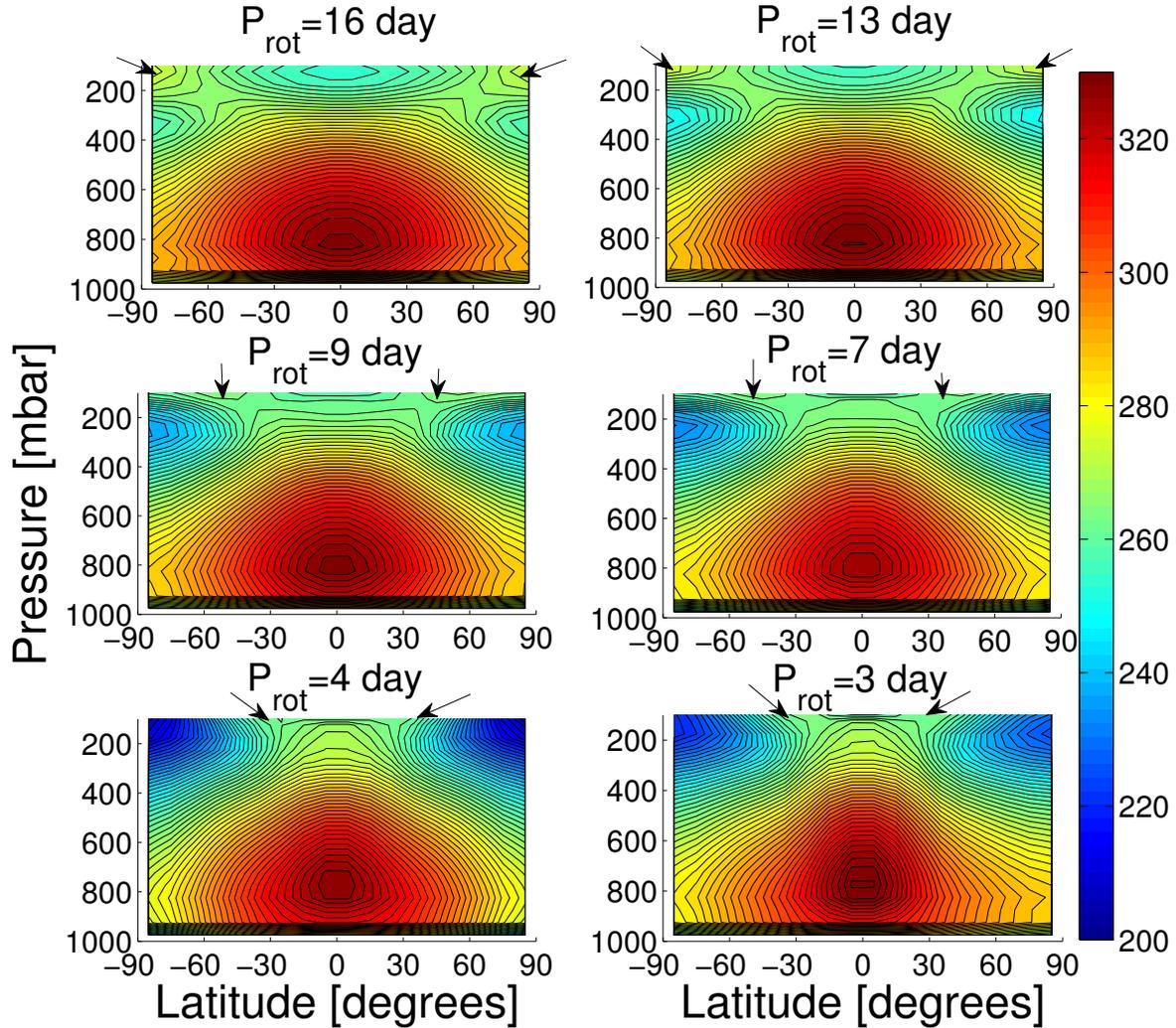}
\caption{Zonally and temporally (1000~days) averaged temperature in K for a 2 Earth size planet around the $L_R/R_P=1$ and $\lambda_R/R_P=0.5$ region, that is, for $P_{rot}=16,13,9,7,4$, and 3~days, respectively. Contour intervals are 2~K. Arrows indicate location of thermal inversions.}
\label{fig: T_Prot16d_3d_Rp2}
\end{figure*} 

We will demonstrate, in the following, that two different types of upper atmosphere thermal inversion can be found in our simulation results. One type of thermal inversion develops because of chevrons around the very fast superrotating equatorial jet. The other type was already mentioned in Sections~\ref{sec: lambda_1} and \ref{sec: L/R-1}: Thermal inversions form at the location of the previously identified local temperature maxima over the cyclonic vortices at $p=225$~mbar, which are the result of adiabatic heating of air masses flowing downwards towards region of reduced geopotential height. We found in Section~\ref{sec: circulation}, however, that thermal inversion should only be present for Rossby climate state regimes between $\lambda_R/R_P \leq 0.5$ and $\lambda \leq 1$.

When inspecting the horizontal cross section at the top of the atmosphere for very fast rotating planets with $P_{rot}=1$~days (Figure~\ref{fig: p225mb_Prot1d}), it becomes indeed clear that the temperature inversion can be linked to the 'chevron' that the superrotating equatorial jet develops, in particular, on larger planets. The 'chevron' for the fast rotating large planets links our study to atmosphere models of hot Jupiters that also show this 'chevron'  shape (e.g. \cite{Kataria2015}). The upper atmosphere temperature inversions associated with the chevron does not change in size and altitude with larger planet size but expand horizontally towards higher latitudes.

We find correspondingly the second type of temperature inversions for slower rotations (Figure~\ref{fig: T_Prot16d}). We could also indeed verify that they start to develop at the $\lambda_R/R_P=1$ transition region and are not present for very slow rotation with $P_{rot}>> 45$~days (not shown). These type of temperature inversions drop in altitude and increase in vertical extent for larger planet sizes, because of the steeper isentrope slopes in the mid-atmosphere ($p=525$~mbar) below the cyclonic vortices, in which the upper atmosphere flow gets adiabatically heated. The upper atmosphere flow can 'fall' deeper and farther towards the poles.

Furthermore, we expect these temperature inversions to vanish, when the rotation period approaches the $L_R/R_P=1$ and $\lambda_R/R_P=0.5$ transition region (Figure~\ref{fig: T_Prot16d_3d_Rp2}), at the same time as the adiabatic heating over the vortices diminishes and disappears (Section~\ref{sec: L/R-1}).

Indeed, we can show that the temperature inversions reduce in vertical extent and are shifted towards the very top of the atmosphere for faster rotation, despite the steepening of the isentropes that should favour adiabatic heating and thus increase adiabatic warming of the upper air flow. We focus again on simulations for a large planet ($R_P=2 R_{Earth}$) with $P_{rot}=3-16$~days (Figure~\ref{fig: T_Prot16d_3d_Rp2}). The shift towards the sponge layer explains why we saw the thermal maxima disappearing already at $P_{rot}=9$~days, when we inspected the horizontal temperature map at $p=225$~mbar (Figure~\ref{fig: p225mb_L1_Rp2}). 

The weakening of the temperature inversion can again be explained by suppressed upwelling due to circulation cell fragmentation (Figure~\ref{fig: circulation_2Rp}, top and middle panels). For $P_{rot}=4$~day, the feature disappears completely. With even faster rotation, the chevron shape develops and forms a new temperature inversions. It appears first at the equator and extends towards higher latitudes with faster rotation (Figure~\ref{fig: T_Prot16d_3d_Rp2}, lower panels and Figure~\ref{fig: T_Prot1d}). We have to caution that the equatorial jet, the chevron and thus the associated thermal inversion are very close to the sponger layer. Furthermore, we don't resolve the upper atmosphere very well due to our strong focus on the investigation of dynamics in the troposphere. It will have to be investigated, if this type of thermal inversion extends to higher altitudes in a model with a better upper atmosphere resolution.

\section{Observational discrimination between different climate states}

It has been shown that for $L_R/R_P\leq 1$ and $\lambda_R/R_P\leq 0.5$ different climate states are possible for the same set of parameters. These degeneracies in climate states should be taken into account at least for tidally locked terrestrial planets at the inner edge of the habitable zone around M dwarfs (Section~\ref{sec: planet parameters}). We will now highlight several simulation results that provide insights into the observational discrimination between different climate states.

In a climate state dominated by a standing tropical Rossby wave, the hottest point is displaced eastward with respect to the substellar point by strong equatorial superrotation as already predicted by \cite{Showman2002}. We find for the $P_{rot}=1$~day simulations that the hottest point is indeed shifted eastward - as expected (Figure~\ref{fig: p225mb_Prot1d}). The eastward displacement of the hot atmosphere region was inferred for the transiting hot Jupiter HD~189733~b from Spitzer lighcurves. The orbital phase variations showed that the maximum of the planet's phase curve and thus the maximum in planetary thermal emission occurred \textbf{prior} to the secondary eclipse \citep{Knutson2007,Knutson2009}.

The hottest point is, however, not shifted eastward but \textbf{westward} with respect to the substellar point in climate states with a strong extra tropical Rossby wave component (Figure~\ref{fig: p225mb_L05_Rp2} top and middle panel and Figure~\ref{fig: p225mb_Prot3d}, left panel). Such a climate state would thus show up in the planet's orbital phase curve because the maximum would occur \textbf{after} the secondary eclipse.

The $P_{rot}=3$~days simulation for $R_P=1.25 R_{Earth}$ shows another mixed climate state, one in which standing tropical and extra tropical Rossby waves contribute in approximately equal measures (Figure \ref{fig: Prot3d_dyn}, left). The result is a hot region eastward \textbf{and} westward from the substellar point (Figure~\ref{fig: p225mb_Prot3d}, left). The temperature map is reminiscent of the results of \cite{Kataria2014} and \cite{Lewis2010}, who also find mixed states with high latitude jets and a relative strong equatorial superrotation in simulations with solar metallicity for the Super-Earth GJ~1214b and hot Neptune GJ~436b. They assumed, however, gas giant composition, that is, the absence of a solid surface. A climate state with a smeared out hot atmosphere region around the substellar point would show up as a broadening of the planets thermal emission around secondary eclipse \citep{Kataria2015}.

The relative low atmosphere temperature ($T\approx 300$~K) and small horizontal temperature differences ($\Delta T_{max}\approx 50$~K) discussed in this study, preclude currently the observational identification of different climate states for a transiting habitable planet. It may be possible, however, to discriminate between different climate states for hot Super-Earths. Future work will, therefore, extend this study to scenarios with stronger thermal forcing.

Interestingly, the Kepler orbital phase curve of the hot Jupiter Kepler~7b shows indeed westward displacement of the maximum in planetary flux with respect to the substellar point \citep{Demory2013}. Thermal emission was ruled out by the authors because superrotation and thus an eastward displacement of the hottest atmosphere was expected. The westward displacement was instead attributed to reflected stellar light from inhomogeneous planet cloud coverage. It is too early to infer that Kepler~7b may be in a climate state with a strong extra tropical Rossby wave component. It would also be a puzzling result, because extra tropical Rossby waves are only expected for small planets with a solid surface \citep{Showman2011}. We find, indeed, the strongest signals of extra tropical Rossby wave components for the smallest planets in our study (Figures \ref{fig: Prot3d_dyn} and~\ref{fig: p225mb_Prot3d} with $R_P=1 R_{Earth}$ and $1.25 R_{Earth}$).

On the other hand, \cite{Kataria2015} showed that hot Jupiter climate models that impose large frictional drag exhibit an equatorial superrotating jet and two additional westerly jets at mid-latitude - just like in the climate states with a mixture of tropical and extra tropical Rossby waves identified in this study. Thus, large Super Earths with a solid surface boundary \citep{Zalucha2013} and gas planets that experience frictional drag inside their atmospheres \citep{Rauscher2013,Kataria2015} may show similar climate state degeneracies. In an upcoming study, we will show that the appearance of a strong extra tropical Rossby wave component in large scale circulation is strongly tied to frictional drag at the surface boundary.

\section{Summary} 
We found that standing tropical and extra tropical Rossby waves are responsible for different climate regime phases on tidally locked habitable planets in this very thorough parameter space investigation that covers $P_{rot}=1-100$~days and $R_P=1- 2 R_{Earth}$. Thus, we could tie climate state phase transitions to transitions in the meridional Rossby wave numbers $\lambda_R/R_P$ and $L_R/R_P$ for the tropical and extratropical Rossby wave number, respectively.

\begin{itemize}
\item \textbf{The formation of a standing tropical Rossby wave at $\lambda_R/R_P=1$} brings about faster zonal wind speeds, which is in line with the explanation of \cite{Showman2011} that the coupling between standing tropical Rossby waves and Kelvin waves transport angular momentum from high to lower latitudes. Furthermore, cyclonic vortices gain in strength and zonal wind maxima at mid-latitudes appear that get more and more confined towards the equator with increasing rotation. The vortices are cold features in the lower and middle atmosphere and thus have a low geopotential height. If upwelling flow over the hot substellar point with large geopotential height 'falls into' these vortices, the flow is adiabatically heated which leads to the formation of temperature inversions in the vertical temperature structure. Furthermore, small embedded reverse circulation cells form inside the direct circulation cells. The change in circulation state leads to a decrease in night side surface temperatures.
\item The $L_R/R_P=1$ transition region and $\lambda_R/R_P=0.5$ transition region almost overlap in rotation period. Because our model favours a climate state dominated by tropical Rossby waves, \textbf{the $\lambda_R/R_P=0.5$ transition is vastly more important in our model}. Once the transition is crossed, zonal wind speed - generally - increases strongly with faster rotation. The transition in tropical Rossby wave number is also associated with beginning fragmentation of the direct circulation cell into a lower and upper segment, which leads to a decrease in upwelling at the top of the atmosphere. As one consequence, the upper atmosphere temperature inversions diminish. At the same time, the lower direct circulation segment allows to warm the night side more efficiently, which leads to an increase in night side surface temperatures.
\item In contrast to our simulation results, \textbf{the $L_R/R_P=1$ transition} is more important in the model of \cite{Edson2011}. The authors report abrupt transition to lower zonal wind speeds. They, furthermore, describe the transition from a climate state with a broad equatorial zonal jet to a state with two zonal jets at higher latitudes. Thus, we conclude that the model of \cite{Edson2011} assumes a second possible climate state in this rotation period region dominated by the extra tropical Rossby wave. 
\item With\textbf{ transition to $L_R/R_P \leq 0.5$}, both the tropical and extra tropical Rossby wave can 'fit' on the planet. And indeed, we find several examples of climate states that show clearly the simultaneous presence of the tropical and extra tropical Rossby waves. Even for $P_{rot}=1$~days, there still exist remnants of a standing extra tropical Rossby wave that leads to broadening of the equatorial jet. In general, however, the tropical Rossby wave dominates even in this climate state regime - at least in our model.
\item The dominance of tropical Rossby waves explains the formation of a single strong equatorial superrotating jet with very fast wind speeds of up to 300~m/s with \textbf{very fast rotations $P_{rot}\leq 1.5$~days}. We have to caution, however, that the exact value of the wind speed should be taken with a grain of salt. The superrotating jet are pushed with faster rotations into our sponge layer, whose friction efficiency needed to be increased to ensure numerical stability. The physical mechanism behind the formation of fast superrotation is, however, already firmly established from the atmosphere modelling of hot Jupiters (e.g. \cite{Kataria2015}). Furthermore, we can explicitly demonstrate the meridional tilt of Rossby wave gyres arising from the standing tropical Rossby wave as proposed by \cite{Showman2011}. Thus, we argue that while we are uncertain about the exact value of the wind speeds, the qualitative assessment about the formation of an equatorial superrotating jet with Rossby wave number $\lambda_R/R_P\leq 0.5$ is sound.
\item The fast equatorial superrotation for $P_{rot}\leq 1.5$~days suppresses the direct equatorial circulation cell practically completely. Circulation is instead provided by the secondary circulation cells whose strength increases with faster rotation. The suppression of the direct circulation is in stark contrast to Earth circulation studies (\cite{Navarra2002}) and that of \cite{Edson2011}, where the direct circulation cell is always the strongest cell. These studies don't show, however, strong equatorial superrotation, which appears to vertically push the already fragmented direct circulation cell by the embedded reverse circulation cells towards the surface. Apparently, superrotation and direct circulation don't mix very well. The suppression of the direct circulation prevents the substellar point at the surface to cool down via upwelling: Thus, the substellar point surface temperatures reach high temperatures. 
\end{itemize}

Furthermore, it was found that larger planets show in general larger wind speeds, steeper isentropes, more pronounced warm upper atmosphere temperature inversions,  warmer substellar point and cooler night side surface temperatures compared to an Earth-size planet with the same rotation period. Furthermore, the largest planet ($R_P=2 R_{Earth}$) exhibits strongly fragmented circulation between $P_{rot}=2-6$~days in contrast to smaller planets. Apparently, circulation and thus potentially cloud formation may be disrupted in this rotation regime. 

In addition, large very fast rotating planets ($R_P=2 R_{Earth}$ and $P_{rot}=1$~day) develop features that are already known from the atmosphere modelling of hot Jupiters, Neptunes and Super-Earths (e.g. \cite{Kataria2015} and \cite{Zalucha2013}): Counter rotating jets at higher latitudes that flank the very fast equatorial superrotating jets, and warm chevrons at the substellar point. Apparently, large planet size and fast rotation are already sufficient for the formation of such features, even though we keep at all times temperate Earth-like forcing. 

\section{Conclusions \& Outlook}

We performed a parameter study with unprecedented thoroughness in planet rotation period ($P_{rot}=1-100$~days) and planet size ($R_P=1- 2 R_{Earth}$) to coherently investigate climate state changes of tidally locked habitable terrestrial planets around M dwarfs.  We were not only capable of identifying more climate state transition states as \cite{Edson2011} and \cite{Merlis2010}, we also confirmed the existence of multiple climate states in the fast rotation regime ($P_{rot}\leq 12$~days for $R_P=2 R_{Earth}$). The existence of multiple climate state solutions was already indicated by \cite{Edson2011}, who speculated about the involvement of the extra tropical Rossby wave. We were, however, capable of identifying the possible formation of either the standing tropical or the extra tropical Rossby waves as the origin of the multiple climate states. Our model apparently favours the first state which exhibits fast equatorial superrotation and the model of \cite{Edson2011} favours apparently the second state with two high to mid-latitude westerly jets with slower wind speeds. 

We attribute the differences between the results from our model and that of \cite{Edson2011} to the relevant strength of the tropical Rossby wave in our model in contrast to the model of \cite{Edson2011}, where the extra tropical Rossby wave dominates.  We will show in an upcoming paper that an increase of surface friction efficiency in our model can also produce climates that are dominated by extra tropical Rossby waves like in \cite{Edson2011}. Reduction of surface friction time scales, on the other hand, increases the strength of tropical Rossby waves and leads to even faster equatorial superrotation. The coherent discussion of climate state regimes and Rossby wave properties allowed us, thus, to identify sources of discrepancies between different models. 

We found, in addition, another region of climate state degeneracies that \cite{Edson2011} missed: For very fast rotations ($P_{rot}\leq 12$~days for $R_P=2 R_{Earth}$), a third possible climate state solution appears, because the tropical and extra tropical Rossby wave can both form standing waves and allow for the existence of a mixed climate state with consequently three zonal wind jets. 

Interestingly, the upper atmosphere's hot spot is eastward shifted with respect to the substellar point in the first state, westward shifted in the second state and the third state shows a longitudinal 'smearing' of the spot across the substellar point. Such climate states could potentially be observed in the orbital phase curve of transiting planet, because the maximum would occur prior, after or 'smeared around' the secondary eclipse. The relative low atmosphere temperature ($T\approx 300$~K) and small horizontal temperature differences ($\Delta T_{max}\approx 50$~K), however, make the observational identification of different climate states for a transiting habitable planet currently infeasible. It may be possible, however, to discriminate between different climate states for hot Super-Earths from the planet orbital phase shift with respect to the secondary eclipse. Future work will, therefore, extend this study to scenarios with stronger thermal forcing.

How much are the investigated climate states of relevance for habitable planets? The planet sizes between $1 R_{Earth}$ and $2 R_{Earth}$ are indeed relevant as this planet mass and size range is compatible with rocky terrestrial Super-Earth planets \citep{Madhusudhan2012}. The relevance of atmosphere dynamics for short orbital periods $P_{rot}\leq 8$~days is more difficult to assess. The conventional definition of the habitable zone (\cite{Kasting1993},\cite{Koppa2013}) puts the inner edge of the habitable zone for the smallest M dwarf star ($M_*=0.2 M_{Sun}$) at 0.08~AU which corresponds to a rotation period of 18~days for the fastest rotating habitable planet. \cite{Yang2014} showed that the habitable zone can be expanded towards the star due to cloud formation at the substellar point, which increases the albedo. Furthermore, \cite{Zsom2013} showed that the habitable zone can be pushed even closer to the host star, if the terrestrial planet is assumed to be relatively dry. The inner edge is located at 0.04~AU for $M_*=0.2 M_{Sun}$, which corresponds to $P_{rot}=$6.5~days. With this new definition of the inner habitable zone, our results become very relevant: A large $R_P=2 R_{Earth}$ planet may, for example, exhibit fragmented circulation cells for $P_{rot}=6$~days, which raises immediately the question, if water clouds can form.

The very fast rotation regime $P_{rot}\leq 1.5$~days provides, in any case, already a link towards the study of atmosphere dynamics on gas planets: We have shown that the largest planet in our study develops fast equatorial superrotation and a temperature chevron that are atmosphere dynamics features well known from hot Jupiters. At the same time, even our largest planet still has remnants of extra tropical Rossby waves that are lacking on hot Jupiters due to their lack of a solid boundary \citep{Showman2011}. Thus, our study provides already at this stage insights about the transition of climate atmosphere dynamics from small habitable terrestrial planets to larger Mini-Neptunes. This transition region is of great interest for Super-Earth planets that show no clear-cut distinction between rocky and gas planets (e.g. \cite{Madhusudhan2012},\cite{Rogers2010}).

As already mentioned, we will strengthen in upcoming follow-up studies the connection between extra tropical Rossby waves and the presence of a solid boundary. The westward displacement of the upper atmosphere's hot spot with respect to the substellar point may then reveal the presence of a solid surface or other sources of frictional drag. 

\section*{Acknowledgments}
We acknowledge support from the KU Leuven projects IDO/10/2013 and GOA/2015-014 (2014–2018 KU Leuven). The computational resources and services used in this work were provided by the VSC (Flemish Supercomputer Center), funded by the Hercules Foundation and the Flemish Government – department EWI. L.C. would further like to thank Olivia Venot for useful comments on the paper. We thank the anonymous referee, whose insightful remarks helped to greatly improve the quality of this paper.

\bibliography{GCM} % your references Yourfile.bib
% Stop your text
\label{lastpage}
\end{document}